\documentclass[a4paper,onecolumn,superscriptaddress,11pt,accepted=2019-04-04]{quantumarticle}

\usepackage[numbers,sort&compress]{natbib}
\usepackage{hyperref}

\usepackage{amsmath,amsthm,amssymb}
\usepackage{xspace,enumerate,color,epsfig}
\usepackage{graphicx}
\graphicspath{{.}{./figures/}}
\usepackage{marvosym}
\usepackage{bm}

\usepackage{stmaryrd}
\usepackage{docmute}
\usepackage{keycommand}

\usepackage{tikzit}

\tikzstyle{white dot}=[inner sep=0mm, minimum size=2mm, draw=black, shape=circle, draw=black, fill=white]
\tikzstyle{gray dot}=[inner sep=0mm, minimum size=2mm, draw=black, shape=circle, draw=black, fill={rgb,255: red,191; green,191; blue,191}]
\tikzstyle{white phase dot}=[minimum size=5mm, font={\footnotesize}, shape=rectangle, rounded corners=2mm, inner sep=0.2mm, outer sep=-2mm, scale=0.8, tikzit shape=circle, draw=black, fill=white, tikzit draw=blue]
\tikzstyle{gray phase dot}=[minimum size=5mm, font={\footnotesize}, shape=rectangle, rounded corners=2mm, inner sep=0.2mm, outer sep=-2mm, scale=0.8, tikzit shape=circle, draw=black, fill={rgb,255: red,191; green,191; blue,191}, tikzit draw=blue]
\tikzstyle{small hadamard}=[fill=white, draw, inner sep=0.6mm, minimum height=1.5mm, minimum width=1.5mm, tikzit shape=rectangle]
\tikzstyle{small dot}=[inner sep=0.7mm, minimum width=0pt, minimum height=0pt, draw, shape=circle]
\tikzstyle{small white dot}=[small dot, fill=white]
\tikzstyle{small black dot}=[small dot, fill=black]
\tikzstyle{special dot}=[small white dot]
\tikzstyle{mbqc dot}=[small black dot]
\tikzstyle{mbqc input dot}=[small white dot]
\tikzstyle{mbqc output dot}=[small gray dot]
\tikzstyle{label}=[font={\footnotesize}, text height=1.5ex, text depth=0.25ex, yshift=0.5mm]
\tikzstyle{left label}=[label, anchor=east, xshift=1.5mm]
\tikzstyle{right label}=[label, anchor=west, xshift=-1.5mm]
\tikzstyle{inline text}=[text height=1.5ex, text depth=0.25ex, yshift=0.5mm]
\tikzstyle{small box}=[rectangle, inline text, fill=white, draw, minimum height=5mm, yshift=-0.5mm, minimum width=5mm, font={\small}]
\tikzstyle{medium box}=[rectangle, inline text, fill=white, draw, minimum height=5mm, yshift=-0.5mm, minimum width=10mm, font={\small}]

\tikzstyle{simple}=[-]
\tikzstyle{gs edge}=[-]
\tikzstyle{gs double edge}=[-, double, shorten <=-1mm, shorten >=-1mm, double distance=2pt]
\tikzstyle{gray edge}=[-, {gray!40!white}]

\let\olddagger\dagger
\renewcommand{\dagger}{\ensuremath{\olddagger}\xspace}

\newcommand{\bra}[1]{\ensuremath{\left\langle #1 \right|}}
\newcommand{\ket}[1]{\ensuremath{\left|  #1 \right\rangle}}

\newcommand{\ketbra}[2]{\ensuremath{\ket{#1}\!\bra{#2}}}

\theoremstyle{definition}

\newtheorem*{remark*}{Remark}
\newtheorem*{convention*}{Convention}

\usepackage[color]{changebar}

\usepackage{color}
\def\bR{\begin{color}{red}} 
\def\bB{\begin{color}{blue}}
\def\bM{\begin{color}{magenta}}
\def\bC{\begin{color}{cyan}}
\def\bW{\begin{color}{white}}
\def\bBl{\begin{color}{black}} 
\def\bG{\begin{color}{green}}
\def\bY{\begin{color}{yellow}}
\def\e{\end{color}\xspace}
\newcommand{\bit}{\begin{itemize}}
\newcommand{\eit}{\end{itemize}\par\noindent}
\newcommand{\ben}{\begin{enumerate}}
\newcommand{\een}{\end{enumerate}\par\noindent}
\newcommand{\beq}{\begin{equation}}
\newcommand{\eeq}{\end{equation}\par\noindent}
\newcommand{\beqa}{\begin{eqnarray*}}
\newcommand{\eeqa}{\end{eqnarray*}\par\noindent}
\newcommand{\beqn}{\begin{eqnarray}}
\newcommand{\eeqn}{\end{eqnarray}\par\noindent}

\input{dots.tikzdefs}

\begin{document}
\title[MQBC with parity-phase interactions]{Universal MBQC with generalised parity-phase interactions and Pauli measurements}

\author{Aleks Kissinger}
\affiliation{Radboud University Nijmegen}
\email{aleks@cs.ru.nl}
\homepage{https://www.cs.ru.nl/A.Kissinger}
\orcid{0000-0002-6090-9684}

\author{John van de Wetering}
\affiliation{Radboud University Nijmegen}
\email{john@vdwetering.name}
\homepage{http://vdwetering.name}
\orcid{0000-0002-5405-8959}


\begin{abstract}
  We introduce a new family of models for measurement-based quantum computation which are deterministic and approximately universal. The resource states which play the role of graph states are prepared via 2-qubit gates of the form $\exp(-i\frac{\pi}{2^{n}} Z\otimes Z)$. When $n = 2$, these are equivalent, up to local Clifford unitaries, to graph states. However, when $n > 2$, their behaviour diverges in two important ways. First, multiple applications of the entangling gate to a single pair of qubits produces non-trivial entanglement, and hence multiple parallel edges between nodes play an important role in these generalised graph states. Second, such a state can be used to realise deterministic, approximately universal computation using only Pauli $Z$ and $X$ measurements and feed-forward. Even though, for $n > 2$, the relevant resource states are no longer stabiliser states, they admit a straightforward, graphical representation using the ZX-calculus. Using this representation, we are able to provide a simple, graphical proof of universality. We furthermore show that for every $n > 2$ this family is capable of producing all Clifford gates and all diagonal gates in the $n$-th level of the Clifford hierarchy.
\end{abstract}

\noindent{\it Keywords}: Measurement-Based Quantum Computation, ZX-calculus, Clifford hierarchy

\section{Introduction}

Measurement-based quantum computation (MBQC) describes a family of alternatives to the quantum circuit model, where one starts with a highly entangled resource state and performs all computation via measurements. 
While individual measurements introduce non-determinism, a deterministic quantum computation can be recovered by allowing future operations to depend on past measurement outcomes, a concept known as \textit{feed-forward}. 
The most prevalent model of MBQC is the \textit{one-way model}, first described by Briegel and Raussendorf \cite{MBQC1}, which makes use of \textit{cluster states} or more generally \textit{graph states} and single-qubit measurements.
Graph states are a family of stabiliser states that can be described by an undirected graph, where each edge represents a pair of qubits entangled via a controlled-$Z$ operation. Since they are stabiliser states, any computation involving just Pauli measurements will be efficiently classically simulable, due to the Gottesman-Knill theorem~\cite{aaronsongottesman2004}. In fact, any computation involving Pauli measurements can be done in a single time-step \cite{MBQC2}. Hence, the entire computational power of the one-way model comes from introducing non-stabiliser measurements to a computation.

It is natural to ask if we can invert this problem: is it possible to obtain universal computation by means of a non-stabiliser resource state and just Pauli measurements? 
There are several ways to achieve this. For example, one could consider resource states which are prepared just like cluster states, but with certain qubits prepared in a $\ket{T}$ magic state rather than the usual $\ket{+}$ state. 
One can also consider \textit{hypergraph states}~\cite{gachechiladze2018changing}, a generalisation of graph states produced by multi-qubit $n$-controlled-$Z$ operations, represented graphically as hyper-edges. These were recently shown to admit a universal model of computation using Pauli measurements and feed-forward~\cite{takeuchi2018quantum}. A different approach was taken in \cite{miller2016hierarchy}, where a resource state was created that allowed non-deterministic approximately universal computation using just X, Y and Z measurements.

In this paper, we introduce a new family of generalisations of graph states which admit universal deterministic computation using only Pauli X and Z measurements and feed-forward. We call these \textit{parity-phase} graph states, or P-graph states. Edges in P-graph states represent an application of the following \textit{parity-phase gate}, for some fixed angle $\alpha$:
\[\hfill P(\alpha) = \exp(-i\frac\alpha2 Z\otimes Z) \hfill\]
We refer to this as a parity-phase gate because it introduces a relative phase of $\alpha$ between its even-parity eigenstates $\ket{00}, \ket{11}$ and its odd parity eigenstates $\ket{01}, \ket{10}$.

We focus on parity-phase gates because they are a popular primitive two-qubit entangling gate in hardware implementations of quantum computation, such as ion trap qubits (via M\o{}lmer S\o{}rensen gates~\cite{iontrapoxford,iontrapcolorado}) and, in transmon-based superconducting qubits \cite{superconductingdelft}. At the end of Section~\ref{sec:universal}, we comment briefly on near-future prospects of implementing this scheme using the latter.

When $\alpha = \frac\pi2$, we obtain resource states which are equivalent to graph states from the one-way model up to local Clifford operations. However, if $\alpha = \frac\pi4$, we can construct resources which are approximately universal for quantum computation using only single-qubit Pauli $X$ and $Z$ measurements and feed-forward. We call this the PPM model, for \textit{parity-phase with Pauli measurements}.\footnote{Note that, in a previous version of this article, this model was referred to as MSPM, for \textit{M\o{}lmer-S\o{}rensen with Pauli measurments}.}

A key feature which distinguishes P-graph states from standard graph states is that, unlike for controlled-Z gates, $P(\alpha)P(\alpha) \neq 1$, except in the degenerate case where $\alpha = \pi$. Hence it is possible, and even desirable, to consider resource states described by graphs with multiple, parallel edges between nodes, e.g.
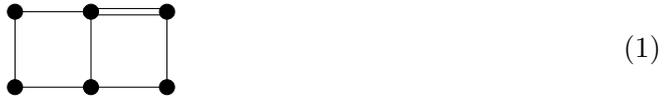
\begin{equation}\label{eq:first-ex}
	\begin{tikzpicture}
	\begin{pgfonlayer}{nodelayer}
		\node [style=small black dot] (0) at (-2, 1) {};
		\node [style=small black dot] (1) at (0, 1) {};
		\node [style=small black dot] (2) at (-2, -1) {};
		\node [style=small black dot] (3) at (0, -1) {};
		\node [style=small black dot] (4) at (2, 1) {};
		\node [style=small black dot] (5) at (2, -1) {};
	\end{pgfonlayer}
	\begin{pgfonlayer}{edgelayer}
		\draw [style=gs edge] (0.center) to (1.center);
		\draw [style=gs edge] (1.center) to (3.center);
		\draw [style=gs edge] (3.center) to (2.center);
		\draw [style=gs edge] (2.center) to (0.center);
		\draw [style=gs edge] (5.center) to (4.center);
		\draw [style=gs edge] (3.center) to (5.center);
		\draw [style=gs double edge] (1.center) to (4.center);
	\end{pgfonlayer}
\end{tikzpicture}
\end{equation}
These parallel edges correspond to multiple applications of the entangling gate $P(\alpha)$ to the adjacent qubits. For example, a doubled edge above indicates the application of $P(\alpha)^2 = P(2\alpha)$. In the graph theory literature, graphs such as \eqref{eq:first-ex} are sometimes referred to as \textit{undirected multigraphs}.

In the PPM model, doubled edges play a special role. Since $P(\frac\pi4)^2 = P(\frac\pi2)$ is equivalent, up to local Clifford operations, to a controlled-Z gate, subgraphs of a P-graph state containing only doubled edges behave in much the same way as traditional graph states. However, P-graph states additionally yield the ability to selectively inject $\pi/4$ phases into computations via nodes connected by single edges. One way to conceptualise this fact is to consider the two-qubit gates $P(\pi/4)$ as introducing `virtual' magic states between pairs of qubits. The phase data carried by this `virtual' magic state can either be destroyed or injected on to one of the neighbouring qubits, depending on the measurement choices, using a method similar to e.g.~Ref.~\cite{bravyi2005universal}.

Notably, this dichotomy gives a clean separation of the efficiently simulable parts of the computation and the rest. In deriving a universal scheme for computation with P-graph states, we will note the feed-forward is only required in the vicinity of single edges. So, much like the case in the one-way model, the entire `Clifford' part of the computation can be done in a single time step.

We make use of the \textit{ZX-calculus}~\cite{CD2} for representing and reasoning about P-graph states. This is a formalism for representing quantum states as certain tensor networks, called ZX-diagrams, as well as a set of rules for transforming and simplifying ZX-diagrams. In some sense, it can be regarded as an enhanced version of the stabiliser formalism, in that any stabiliser state (or Clifford circuit) can be described as a ZX-diagram, and if two such states/circuits are equal, then one can be efficiently transformed into the other by means of the ZX-calculus~\cite{Backens1}.
However, the ZX-calculus goes beyond the stabiliser formalism in that arbitrary linear maps between qubits can be described as ZX-diagrams. It was recently shown that two ZX-diagrams describing the same linear map can always be transformed into one another using an extended version of the ZX-calculus (albeit not necessarily efficiently)~\cite{SimonCompleteness,HarnyCompleteness}. Our usage of the calculus lies somewhere between these two extremes: we will use the ZX-calculus to efficiently manipulate certain well-behaved families of non-stabiliser states. 

In particular, we use a variation of the translation from graph states and measurement patterns into ZX-diagrams given by Duncan and Perdrix \cite{DP2} to prove the correctness of our computation scheme. Much like in their work, and in standard approaches to MBQC, we will demonstrate the possibility of deterministic computation by `pushing' Pauli errors forward from measurements to qubits in future, which can be corrected. However, unlike previous work, we rely on the extra flexibility of ZX-diagrams to represent non-stabilizer correlations between qubits and develop techniques for `pushing' errors through these edges using the diagrammatic language. As we shall see, the diagrams keep track of the extra (Clifford) errors introduced by propagating errors forward, and it enables us to derive a technique for performing Pauli and Clifford corrections purely by means of single-qubit measurement choices in the bases $\{ X, Z \}$.
This yields a measurement-based model which is very flexible. To give some evidence of this flexibility, we show in Section~\ref{sec:clifhier} how to generalise to P-graph states where a single edge denotes an application of $P(\frac{\pi}{2^{n-1}})$. This enables us to incorporate a familiar `trick' (see e.g.~\cite[Section III]{Gottesman}) into the model to deterministically implement any diagonal gate of the $n$-th level of the Clifford hierarchy.

\textbf{Related work.} To give a proof of universality, we introduce `hairy brickwork states', which are inspired by the brickwork states introduced in Ref.~\cite{brickworkuniversal} for universal computation in the one-way model. Alternatives to the one-way model have been considered, notably in a broad range of models by Gross \textit{et al} \cite{GrossEisartBeyond}, which include a variation on graph states called \textit{weighted graph states}, whose two-qubit interactions are equivalent to $P(\alpha)$ for values of $\alpha$ different from $\pi/2$, up to local unitaries.  However, our approach is distinct in that we use very limited measurements and rely on the fact that P-graph states are a more powerful resource than standard graph states, in spite of having non-maximal entanglement between some pairs of qubits. Our scheme also has the property of the one-way model that all errors can be corrected via feed-forward, eliminating the need for `trial-until-success' strategies used by Ref.~\cite{GrossEisartBeyond}. 
In Ref.~\cite{miller2016hierarchy} a model based on hypergraph states is constructed that only needs Pauli measurements to become universal, but its structure is more complex than ours and the protocols used are not deterministic. 
Deterministic protocols using hypergraph states and Pauli measurements have been proposed since the appearance of this article in preprint, namely those in Refs.~\cite{takeuchi2018quantum} and \cite{gachechiladze2018changing}. However, our protocol remains interesting for several reasons. First, parity-phase interactions are typically more primitive, in that they have simpler realisations within the gate sets of current hardware proposals. Second, the universal gate set we produce in our model is Clifford+T (or more generally, Clifford + arbitrary diagonal Clifford-hierarchy gates), as opposed to CCZ+Hadamard. While the latter is also universal, it requires extra overhead for encoding computations in a higher-dimensional space~\cite{ShiToffoliHadamard}. Third, and perhaps most importantly, we introduce a drastically different methodology to existing approaches. This yields a rather flexible family of models that enable us to explore a variety of multi-qubit interactions and graph topologies. Indeed it is a topic of active research to extend these techniques to hypergraph-based models, where the role of the ZX-calculus is played by the recently-introduced ZH-calculus~\cite{backens2018zhcalculus}.


\section{The PPM Model}
A P-graph state is described by an undirected graph, where we allow multiple edges between vertices. In practice, we will only need to consider two cases: either a single edge or a double edge:
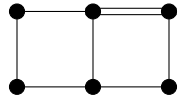
\ctikzfig{ms-graph-state}
A single edge describes the application of an $P(\pi/4)$ gate, whereas a double edge describes the application of an $P(\pi/2) = P(\pi/4)^2$ gate. 

A \textit{measurement pattern} is a P-graph state where each node is labelled by a \textit{measurement expression} of the form ``$b \leftarrow \phi(a_1, \ldots, a_n)$'' where $b$ is a fresh name called the \textit{output value} and $\phi$ is a classical function from boolean variables $a_1, \ldots, a_n$ to a single boolean value. In this case, we say for each $a_i$ that $b$ depends on $a_i$. An MS-pattern is \textit{well-founded} if there are no cyclic dependencies between variables, such as in the following pattern:
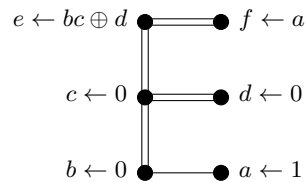
\ctikzfig{ms-pattern}

These expressions do not have any explicit time-ordering, but there are restrictions on the order in which measurements can be made, due to dependencies on prior outcomes, i.e. \textit{feed-forward}. As a matter of convention, we will typically draw earlier measurements below later ones, i.e. `time' flows upward.

Computations are performed as follows:
\begin{enumerate}
  \item A qubit is initialised in the $\ket +$ state for each vertex in a P-graph state.
  \item For every edge in the graph, $P(\frac\pi4)$ is applied to the two qubits at its source and target. In particular, $P(\frac\pi2) = P(\frac\pi4)^2$ is applied to every pair of qubits connected by a double edge.
  \item For a qubit labelled ``$b \leftarrow \phi(a_1, \ldots, a_n)$'', where the values $a_1, \ldots, a_n$ are known, measure in the $X$-basis if $\phi(a_1, \ldots, a_n) = 0$ and the $Z$-basis otherwise. In either case, store the measurement result in $b$.
  \item Optionally, perform some classical post-processing on the measurement results.
\end{enumerate}
As with the one-way model, the two-qubit gates all commute, so the order of application is irrelevant, leaving only the undirected graph structure.

To show that this model is universal, we will compose smaller patterns into larger ones. In order to do this, we give a notion of \textit{pattern fragment} analogous to the notion given for the one-way model. A pattern fragment is just like a measurement pattern, with the exception that we additionally identify two (not necessarily disjoint) subsets of vertices $I, O \subseteq V$ which respectively correspond to inputs and outputs. Inputs correspond to qubits that can be in an arbitrary state, rather than the fixed state $\ket{+}$. Outputs correspond to qubits which remain unmeasured after the application of the pattern-fragment.

Each vertex in $I$ is labelled with an \textit{input error expression} of the form: ``$(z, x) \leftarrow \square$'' for fresh variables $z$ and $x$, which capture whether a Z or X error is being fed forward into this vertex. Unless the input vertex is also an output, the vertex will also be labelled with a measurement expression.
Measurement choice functions $\phi$ in the fragment are allowed to depend on the input errors as well as other measurements occurring within the fragment.

Each vertex in the output set $O$ is labelled by an expression of the form: ``$\square \leftarrow (\zeta, \xi)$'' consisting of a pair of classical functions $\zeta, \xi$ which again can depend on the input errors and the results of measurements in the pattern fragment. Vertices in $O$ are not measured so they do not contain a measurement expression. 

Here is an example of a pattern fragment where the bottom left qubit is an input, and the top left qubit is an output:
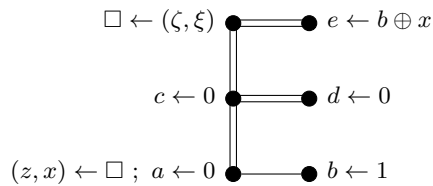
\ctikzfig{ms-pattern-fragment}
We say a pattern fragment implements a gate $G$ if, for any input state of the form: $X^x Z^z \ket\psi$, performing the pattern fragment yields $X^\xi Z^\zeta G \ket\psi$. This extends in the natural way to gates with multiple input/output qubits:
\[\hfill
(X^{x_1}Z^{z_1} \otimes \ldots \otimes X^{x_n}Z^{z_n}) \ket\psi
\ \ \mapsto\ \ 
(X^{\xi_1}Z^{\zeta_1} \otimes \ldots \otimes X^{\xi_m}Z^{\zeta_m}) G \ket\psi
\hfill\]

Composing pattern fragments then results in the composition of their associated gates, hence it suffices for the sake of universality to show we can implement a universal set of quantum gates via pattern fragments.

\section{ZX-notation}

To derive a universal set of measurement patterns for the PPM model, it is convenient to use \textit{ZX-notation}, which is a superset of the usual quantum circuit notation. It is used to depict linear maps from qubits to qubits, put together in the usual way via composition and tensor product. We will provide a brief overview. For an in-depth reference see Ref.~\cite{CKbook}.

The only operations in ZX-notation are swap gates, identities, and special linear maps called \textit{spiders}. These come in two varieties, Z-spiders depicted as white dots:
\[\hfill \begin{tikzpicture}
	\begin{pgfonlayer}{nodelayer}
		\node [style=white phase dot] (0) at (0, 0) {$\alpha$};
		\node [style=none] (1) at (1.25, 1) {};
		\node [style=none] (2) at (-1.25, 1) {};
		\node [style=none] (3) at (-1.25, -1) {};
		\node [style=none] (4) at (1.25, -1) {};
		\node [style=none] (5) at (1.25, 0.5) {};
		\node [style=none] (6) at (-1.25, 0.5) {};
		\node [style=none, rotate=90] (7) at (-1, -0.25) {...};
		\node [style=none, rotate=90] (8) at (1, -0.25) {...};
	\end{pgfonlayer}
	\begin{pgfonlayer}{edgelayer}
		\draw [style=simple, in=-141, out=0, looseness=0.75] (3.center) to (0);
		\draw [style=simple, in=180, out=-39, looseness=0.75] (0) to (4.center);
		\draw [style=simple, in=180, out=22, looseness=0.75] (0) to (5.center);
		\draw [style=simple, in=180, out=39, looseness=0.75] (0) to (1.center);
		\draw [style=simple, in=0, out=158, looseness=0.75] (0) to (6.center);
		\draw [style=simple, in=141, out=0, looseness=0.75] (2.center) to (0);
	\end{pgfonlayer}
\end{tikzpicture} \ \ :=\ \ \ketbra{0 \cdots 0}{0 \cdots 0} + e^{i \alpha} \ketbra{1 \cdots 1}{1 \cdots 1} \hfill\]
and X-spiders depicted as grey dots:
\[\hfill \begin{tikzpicture}
	\begin{pgfonlayer}{nodelayer}
		\node [style=gray phase dot] (0) at (0, 0) {$\alpha$};
		\node [style=none] (1) at (1.25, 1) {};
		\node [style=none] (2) at (-1.25, 1) {};
		\node [style=none] (3) at (-1.25, -1) {};
		\node [style=none] (4) at (1.25, -1) {};
		\node [style=none] (5) at (1.25, 0.5) {};
		\node [style=none] (6) at (-1.25, 0.5) {};
		\node [style=none, rotate=90] (7) at (-1, -0.25) {...};
		\node [style=none, rotate=90] (8) at (1, -0.25) {...};
	\end{pgfonlayer}
	\begin{pgfonlayer}{edgelayer}
		\draw [style=simple, in=-141, out=0, looseness=0.75] (3.center) to (0);
		\draw [style=simple, in=180, out=-39, looseness=0.75] (0) to (4.center);
		\draw [style=simple, in=180, out=22, looseness=0.75] (0) to (5.center);
		\draw [style=simple, in=180, out=39, looseness=0.75] (0) to (1.center);
		\draw [style=simple, in=0, out=158, looseness=0.75] (0) to (6.center);
		\draw [style=simple, in=141, out=0, looseness=0.75] (2.center) to (0);
	\end{pgfonlayer}
\end{tikzpicture} \ \ :=\ \ \ketbra{+ \cdots +}{+ \cdots +} + e^{i \alpha} \ketbra{- \cdots -}{- \cdots -} \hfill\]

A special case is when spiders have a single input and output, in which case they form Z-phase and X-phase gates (up to a global phase):
\[\hfill \begin{tikzpicture}
	\begin{pgfonlayer}{nodelayer}
		\node [style=white phase dot] (0) at (0, 0) {$\alpha$};
		\node [style=none] (1) at (-1.25, 0) {};
		\node [style=none] (2) at (1.25, 0) {};
	\end{pgfonlayer}
	\begin{pgfonlayer}{edgelayer}
		\draw [style=simple] (1.center) to (0);
		\draw [style=simple] (0) to (2.center);
	\end{pgfonlayer}
\end{tikzpicture} \ \ =\ \ \ketbra{0}{0} + e^{i \alpha} \ketbra{1}{1} = Z(\alpha) \hfill\]
\[\hfill \begin{tikzpicture}
	\begin{pgfonlayer}{nodelayer}
		\node [style=gray phase dot] (0) at (0, 0) {$\alpha$};
		\node [style=none] (1) at (-1.25, 0) {};
		\node [style=none] (2) at (1.25, 0) {};
	\end{pgfonlayer}
	\begin{pgfonlayer}{edgelayer}
		\draw [style=simple] (1.center) to (0);
		\draw [style=simple] (0) to (2.center);
	\end{pgfonlayer}
\end{tikzpicture} \ \ =\ \ \ketbra{+}{+} + e^{i \alpha} \ketbra{-}{-} = X(\alpha) \hfill\]
Note that we adopt the convention for $\alpha$ such that the Pauli $Z$ and $X$ gates are obtained as $Z(\pi)$ and $X(\pi)$, respectively. Another special case is when $\alpha = 0$, in which case we omit the label:
\[\hfill \begin{tikzpicture}
	\begin{pgfonlayer}{nodelayer}
		\node [style=white dot] (0) at (0, 0) {};
		\node [style=none] (1) at (1.25, 1) {};
		\node [style=none] (2) at (-1.25, 1) {};
		\node [style=none] (3) at (-1.25, -1) {};
		\node [style=none] (4) at (1.25, -1) {};
		\node [style=none] (5) at (1.25, 0.5) {};
		\node [style=none] (6) at (-1.25, 0.5) {};
		\node [style=none, rotate=90] (7) at (-1, -0.25) {...};
		\node [style=none, rotate=90] (8) at (1, -0.25) {...};
	\end{pgfonlayer}
	\begin{pgfonlayer}{edgelayer}
		\draw [style=simple, in=-141, out=0, looseness=0.75] (3.center) to (0);
		\draw [style=simple, in=180, out=-39, looseness=0.75] (0) to (4.center);
		\draw [style=simple, in=180, out=22, looseness=0.75] (0) to (5.center);
		\draw [style=simple, in=180, out=39, looseness=0.75] (0) to (1.center);
		\draw [style=simple, in=0, out=158, looseness=0.75] (0) to (6.center);
		\draw [style=simple, in=141, out=0, looseness=0.75] (2.center) to (0);
	\end{pgfonlayer}
\end{tikzpicture} \ \ :=\ \ \ketbra{0 \cdots 0}{0 \cdots 0} + \ketbra{1 \cdots 1}{1 \cdots 1} \hfill\]
\[\hfill \begin{tikzpicture}
	\begin{pgfonlayer}{nodelayer}
		\node [style=gray dot] (0) at (0, 0) {};
		\node [style=none] (1) at (1.25, 1) {};
		\node [style=none] (2) at (-1.25, 1) {};
		\node [style=none] (3) at (-1.25, -1) {};
		\node [style=none] (4) at (1.25, -1) {};
		\node [style=none] (5) at (1.25, 0.5) {};
		\node [style=none] (6) at (-1.25, 0.5) {};
		\node [style=none, rotate=90] (7) at (-1, -0.25) {...};
		\node [style=none, rotate=90] (8) at (1, -0.25) {...};
	\end{pgfonlayer}
	\begin{pgfonlayer}{edgelayer}
		\draw [style=simple, in=-141, out=0, looseness=0.75] (3.center) to (0);
		\draw [style=simple, in=180, out=-39, looseness=0.75] (0) to (4.center);
		\draw [style=simple, in=180, out=22, looseness=0.75] (0) to (5.center);
		\draw [style=simple, in=180, out=39, looseness=0.75] (0) to (1.center);
		\draw [style=simple, in=0, out=158, looseness=0.75] (0) to (6.center);
		\draw [style=simple, in=141, out=0, looseness=0.75] (2.center) to (0);
	\end{pgfonlayer}
\end{tikzpicture} \ \ :=\ \ \ketbra{+ \cdots +}{+ \cdots +} + \ketbra{- \cdots -}{- \cdots -} \hfill\]

In particular, when these phaseless spiders have single input and output they are the identity:
\ctikzfig{zx-identity}

Phaseless spiders can be thought of as a generalisation of the GHZ state to a linear map. Indeed the Z-spider with 0 inputs and 3 outputs is the usual GHZ state (up to normalisation):
\[\hfill \begin{tikzpicture}
	\begin{pgfonlayer}{nodelayer}
		\node [style=white dot] (0) at (0, 0) {};
		\node [style=none] (1) at (1, 0.75) {};
		\node [style=none] (2) at (1, -0.75) {};
		\node [style=none] (3) at (1, 0) {};
	\end{pgfonlayer}
	\begin{pgfonlayer}{edgelayer}
		\draw [style=simple, in=180, out=-75, looseness=1.00] (0) to (2.center);
		\draw [style=simple] (0) to (3.center);
		\draw [style=simple, in=180, out=75, looseness=1.00] (0) to (1.center);
	\end{pgfonlayer}
\end{tikzpicture} \ =\ \ket{000} + \ket{111} \hfill\]
Furthermore, the $\ket{0}$ and $\ket{+}$ states are just one-legged spiders:
\begin{equation}\label{eq:spider-basis1}
\hfill
\begin{tikzpicture}
	\begin{pgfonlayer}{nodelayer}
		\node [style=white dot] (0) at (-0.5, 0) {};
		\node [style=none] (1) at (0.5, 0) {};
	\end{pgfonlayer}
	\begin{pgfonlayer}{edgelayer}
		\draw [style=simple] (0) to (1.center);
	\end{pgfonlayer}
\end{tikzpicture}\ \ =\ \ \ket{0} + \ket{1} \ \propto \ket{+}
\qquad\qquad
\begin{tikzpicture}
	\begin{pgfonlayer}{nodelayer}
		\node [style=gray dot] (0) at (-0.5, 0) {};
		\node [style=none] (1) at (0.5, 0) {};
	\end{pgfonlayer}
	\begin{pgfonlayer}{edgelayer}
		\draw [style=simple] (0) to (1.center);
	\end{pgfonlayer}
\end{tikzpicture}\ \ =\ \ \ket{+} + \ket{-} \ \propto \ket{0}
\hfill
\end{equation}
The $\ket{1}$ and $\ket{-}$ states are these states followed by respectively a $Z(\pi)$ and $X(\pi)$ gate:
\[\hfill \begin{tikzpicture}
	\begin{pgfonlayer}{nodelayer}
		\node [style=none] (0) at (-2.25, -0) {};
		\node [style=none] (1) at (1, -0) {};
		\node [style=none] (2) at (-1.5, -0.1) {=};
		\node [style={white dot}] (3) at (-4.25, -0) {};
		\node [style={white phase dot}] (4) at (-3.25, -0) {$\pi$};
		\node [style={white phase dot}] (5) at (0, -0) {$\pi$};
	\end{pgfonlayer}
	\begin{pgfonlayer}{edgelayer}
		\draw [style=simple] (3) to (4);
		\draw [style=simple] (4) to (0.center);
		\draw [style=simple] (5) to (1.center);
	\end{pgfonlayer}
\end{tikzpicture}\ \ =\ \ \ket{0} + e^{i\pi} \ket{1} \ \propto \ket{-} \hfill\]
\[\hfill \begin{tikzpicture}
	\begin{pgfonlayer}{nodelayer}
		\node [style={gray dot}] (0) at (-4.25, 0) {};
		\node [style=none] (1) at (-2.25, 0) {};
		\node [style={gray phase dot}] (2) at (-3.25, 0) {$\pi$};
		\node [style=none] (3) at (0.9999999, 0) {};
		\node [style={gray phase dot}] (4) at (0, 0) {$\pi$};
		\node [style=none] (5) at (-1.5, -0.100001) {=};
	\end{pgfonlayer}
	\begin{pgfonlayer}{edgelayer}
		\draw [style=simple] (0) to (2);
		\draw [style=simple] (2) to (1.center);
		\draw [style=simple] (4) to (3.center);
	\end{pgfonlayer}
\end{tikzpicture}\ \ =\ \ \ket{+} + e^{i\pi} \ket{-} \ \propto \ket{1} \hfill\]
Note that hence forth we will ignore non-zero scalar factors and we will cease to write $\propto$ instead of $=$, as scalar factors will not enter into our calculations.

While spiders are typically non-unitary, and hence not quantum gates, they can be used to construct a universal family of quantum gates. We already saw Z- and X-phase gates, while the CNOT gate can be constructed as follows:
\[
\textit{CNOT}\ :=\ \begin{tikzpicture}
	\begin{pgfonlayer}{nodelayer}
		\node [style=white dot] (0) at (0, 0.5) {};
		\node [style=none] (1) at (1.25, 0.5) {};
		\node [style=gray dot] (2) at (0, -0.5) {};
		\node [style=none] (3) at (-1.25, 0.5) {};
		\node [style=none] (4) at (1.25, -0.5) {};
		\node [style=none] (5) at (-1.25, -0.5) {};
	\end{pgfonlayer}
	\begin{pgfonlayer}{edgelayer}
		\draw [style=simple] (0) to (2);
		\draw [style=simple] (0) to (3.center);
		\draw [style=simple] (0) to (1.center);
		\draw [style=simple] (5.center) to (2);
		\draw [style=simple] (2) to (4.center);
	\end{pgfonlayer}
\end{tikzpicture} \ =\ \begin{tikzpicture}
	\begin{pgfonlayer}{nodelayer}
		\node [style=white dot] (0) at (-0.5, 0.5) {};
		\node [style=none] (1) at (1.5, 1) {};
		\node [style=gray dot] (2) at (0.5, -0.5) {};
		\node [style=none] (3) at (-1.5, 0.5) {};
		\node [style=none] (4) at (1.5, -0.5) {};
		\node [style=none] (5) at (-1.5, -1) {};
	\end{pgfonlayer}
	\begin{pgfonlayer}{edgelayer}
		\draw [style=simple, in=120, out=-75, looseness=1.25] (0) to (2);
		\draw [style=simple] (0) to (3.center);
		\draw [style=simple, in=180, out=45, looseness=1.00] (0) to (1.center);
		\draw [style=simple, in=-135, out=0, looseness=1.00] (5.center) to (2);
		\draw [style=simple] (2) to (4.center);
	\end{pgfonlayer}
\end{tikzpicture} \ =\ \begin{tikzpicture}
	\begin{pgfonlayer}{nodelayer}
		\node [style=white dot] (0) at (0.5, 0.5) {};
		\node [style=none] (1) at (-1.5, 1) {};
		\node [style=gray dot] (2) at (-0.5, -0.5) {};
		\node [style=none] (3) at (1.5, 0.5) {};
		\node [style=none] (4) at (-1.5, -0.5) {};
		\node [style=none] (5) at (1.5, -1) {};
	\end{pgfonlayer}
	\begin{pgfonlayer}{edgelayer}
		\draw [style=simple, in=60, out=-105, looseness=1.25] (0) to (2);
		\draw [style=simple] (0) to (3.center);
		\draw [style=simple, in=0, out=135, looseness=1.00] (0) to (1.center);
		\draw [style=simple, in=-45, out=180, looseness=1.00] (5.center) to (2);
		\draw [style=simple] (2) to (4.center);
	\end{pgfonlayer}
\end{tikzpicture} 
\]
Note that we can always reverse the direction of a wire connecting two spiders without changing the value of the linear map, hence we can write the leftmost diagram above without ambiguity. More generally, diagrams of spiders are less rigid than circuits. Much like diagrammatic depictions of tensor networks (e.g.~Ref.~\cite{Penrose}), any two diagrams of spiders with the same connectivity and ordering of inputs/outputs describe the same linear map.

The power of the ZX-notation comes from the set of rewrite rules associated to it, collectively known as the \textit{ZX-calculus}. The ZX-calculus has a couple of variations. We will only need a minimal set of rules. The first two are the \textit{spider-fusion} rules, which says adjacent spiders of the same colour fuse together, and their phases add:
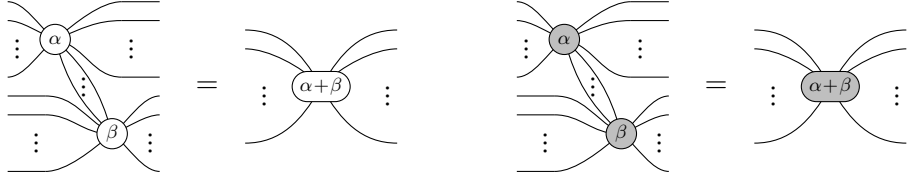
\[\hfill
\begin{tikzpicture}
	\begin{pgfonlayer}{nodelayer}
		\node [style=white phase dot] (0) at (-2.5, -1.25) {$\beta$};
		\node [style=none] (1) at (-1.25, -0.25) {};
		\node [style=none] (2) at (-4.25, -0.25) {};
		\node [style=none] (3) at (-4.25, -2.25) {};
		\node [style=none] (4) at (-1.25, -2.25) {};
		\node [style=none] (5) at (-1.25, -0.75) {};
		\node [style=none] (6) at (-4.25, -0.75) {};
		\node [style=none, rotate=90] (7) at (-4.5, -1.5) {...};
		\node [style=none, rotate=90] (8) at (-1.5, -1.5) {...};
		\node [style=none] (9) at (-5.25, 1.75) {};
		\node [style=white phase dot] (10) at (-4, 1.25) {$\alpha$};
		\node [style=none] (11) at (-2.25, 2.25) {};
		\node [style=none] (12) at (-5.25, 2.25) {};
		\node [style=none] (13) at (-2.25, 0.25) {};
		\node [style=none, rotate=90] (14) at (-2, 1) {...};
		\node [style=none] (15) at (-2.25, 1.75) {};
		\node [style=none] (16) at (-5.25, 0.25) {};
		\node [style=none, rotate=90] (17) at (-5, 1) {...};
		\node [style=none] (18) at (-1.25, 0.25) {};
		\node [style=none] (19) at (-1.25, 2.25) {};
		\node [style=none] (20) at (-1.25, 1.75) {};
		\node [style=none] (21) at (-5.25, -0.75) {};
		\node [style=none] (22) at (-5.25, -2.25) {};
		\node [style=none] (23) at (-5.25, -0.25) {};
		\node [style=none] (24) at (0, 0) {$=$};
		\node [style=none, rotate=90] (25) at (-3.25, 0) {...};
		\node [style=none] (26) at (1, 1.5) {};
		\node [style=none] (27) at (5, 1) {};
		\node [style=none, rotate=90] (28) at (1.5, -0.25) {...};
		\node [style=none] (29) at (5, -1.5) {};
		\node [style=none, rotate=90] (30) at (4.75, -0.25) {...};
		\node [style=none] (31) at (1, 1) {};
		\node [style=white phase dot] (32) at (3, 0) {$\ \alpha\!+\!\beta\ $};
		\node [style=none] (33) at (5, 1.5) {};
		\node [style=none] (34) at (1, -1.5) {};
	\end{pgfonlayer}
	\begin{pgfonlayer}{edgelayer}
		\draw [style=simple, in=-141, out=0, looseness=0.75] (3.center) to (0);
		\draw [style=simple, in=180, out=-39, looseness=0.75] (0) to (4.center);
		\draw [style=simple, in=180, out=22, looseness=0.75] (0) to (5.center);
		\draw [style=simple, in=180, out=39, looseness=0.75] (0) to (1.center);
		\draw [style=simple, in=0, out=180, looseness=1.00] (0) to (6.center);
		\draw [style=simple, in=150, out=0, looseness=1.00] (2.center) to (0);
		\draw [style=simple, in=-141, out=0, looseness=0.75] (16.center) to (10);
		\draw [style=simple, in=180, out=-15, looseness=1.00] (10) to (13.center);
		\draw [style=simple, in=180, out=22, looseness=1.00] (10) to (15.center);
		\draw [style=simple, in=180, out=39, looseness=0.75] (10) to (11.center);
		\draw [style=simple, in=0, out=158, looseness=0.75] (10) to (9.center);
		\draw [style=simple, in=141, out=0, looseness=0.75] (12.center) to (10);
		\draw [style=simple, bend left=15, looseness=1.00] (10) to (0);
		\draw [style=simple] (11.center) to (19.center);
		\draw [style=simple] (15.center) to (20.center);
		\draw [style=simple] (13.center) to (18.center);
		\draw [style=simple] (23.center) to (2.center);
		\draw [style=simple] (21.center) to (6.center);
		\draw [style=simple] (22.center) to (3.center);
		\draw [style=simple, bend left=15, looseness=1.00] (0) to (10);
		\draw [style=simple, in=-120, out=0, looseness=1.00] (34.center) to (32);
		\draw [style=simple, in=180, out=-60, looseness=1.00] (32) to (29.center);
		\draw [style=simple, in=180, out=45, looseness=0.75] (32) to (27.center);
		\draw [style=simple, in=180, out=60, looseness=1.00] (32) to (33.center);
		\draw [style=simple, in=0, out=135, looseness=0.75] (32) to (31.center);
		\draw [style=simple, in=120, out=0, looseness=1.00] (26.center) to (32);
	\end{pgfonlayer}
\end{tikzpicture} \qquad\qquad
\begin{tikzpicture}
	\begin{pgfonlayer}{nodelayer}
		\node [style=gray phase dot] (0) at (-2.5, -1.25) {$\beta$};
		\node [style=none] (1) at (-1.25, -0.25) {};
		\node [style=none] (2) at (-4.25, -0.25) {};
		\node [style=none] (3) at (-4.25, -2.25) {};
		\node [style=none] (4) at (-1.25, -2.25) {};
		\node [style=none] (5) at (-1.25, -0.75) {};
		\node [style=none] (6) at (-4.25, -0.75) {};
		\node [style=none, rotate=90] (7) at (-4.5, -1.5) {...};
		\node [style=none, rotate=90] (8) at (-1.5, -1.5) {...};
		\node [style=none] (9) at (-5.25, 1.75) {};
		\node [style=gray phase dot] (10) at (-4, 1.25) {$\alpha$};
		\node [style=none] (11) at (-2.25, 2.25) {};
		\node [style=none] (12) at (-5.25, 2.25) {};
		\node [style=none] (13) at (-2.25, 0.25) {};
		\node [style=none, rotate=90] (14) at (-2, 1) {...};
		\node [style=none] (15) at (-2.25, 1.75) {};
		\node [style=none] (16) at (-5.25, 0.25) {};
		\node [style=none, rotate=90] (17) at (-5, 1) {...};
		\node [style=none] (18) at (-1.25, 0.25) {};
		\node [style=none] (19) at (-1.25, 2.25) {};
		\node [style=none] (20) at (-1.25, 1.75) {};
		\node [style=none] (21) at (-5.25, -0.75) {};
		\node [style=none] (22) at (-5.25, -2.25) {};
		\node [style=none] (23) at (-5.25, -0.25) {};
		\node [style=none] (24) at (0, 0) {$=$};
		\node [style=none, rotate=90] (25) at (-3.25, 0) {...};
		\node [style=none] (26) at (1, 1.5) {};
		\node [style=none] (27) at (5, 1) {};
		\node [style=none, rotate=90] (28) at (1.5, -0.25) {...};
		\node [style=none] (29) at (5, -1.5) {};
		\node [style=none, rotate=90] (30) at (4.75, -0.25) {...};
		\node [style=none] (31) at (1, 1) {};
		\node [style=gray phase dot] (32) at (3, 0) {$\ \alpha\!+\!\beta\ $};
		\node [style=none] (33) at (5, 1.5) {};
		\node [style=none] (34) at (1, -1.5) {};
	\end{pgfonlayer}
	\begin{pgfonlayer}{edgelayer}
		\draw [style=simple, in=-141, out=0, looseness=0.75] (3.center) to (0);
		\draw [style=simple, in=180, out=-39, looseness=0.75] (0) to (4.center);
		\draw [style=simple, in=180, out=22, looseness=0.75] (0) to (5.center);
		\draw [style=simple, in=180, out=39, looseness=0.75] (0) to (1.center);
		\draw [style=simple, in=0, out=180, looseness=1.00] (0) to (6.center);
		\draw [style=simple, in=150, out=0, looseness=1.00] (2.center) to (0);
		\draw [style=simple, in=-141, out=0, looseness=0.75] (16.center) to (10);
		\draw [style=simple, in=180, out=-15, looseness=1.00] (10) to (13.center);
		\draw [style=simple, in=180, out=22, looseness=1.00] (10) to (15.center);
		\draw [style=simple, in=180, out=39, looseness=0.75] (10) to (11.center);
		\draw [style=simple, in=0, out=158, looseness=0.75] (10) to (9.center);
		\draw [style=simple, in=141, out=0, looseness=0.75] (12.center) to (10);
		\draw [style=simple, bend left=15, looseness=1.00] (10) to (0);
		\draw [style=simple] (11.center) to (19.center);
		\draw [style=simple] (15.center) to (20.center);
		\draw [style=simple] (13.center) to (18.center);
		\draw [style=simple] (23.center) to (2.center);
		\draw [style=simple] (21.center) to (6.center);
		\draw [style=simple] (22.center) to (3.center);
		\draw [style=simple, bend left=15, looseness=1.00] (0) to (10);
		\draw [style=simple, in=-120, out=0, looseness=1.00] (34.center) to (32);
		\draw [style=simple, in=180, out=-60, looseness=1.00] (32) to (29.center);
		\draw [style=simple, in=180, out=45, looseness=0.75] (32) to (27.center);
		\draw [style=simple, in=180, out=60, looseness=1.00] (32) to (33.center);
		\draw [style=simple, in=0, out=135, looseness=0.75] (32) to (31.center);
		\draw [style=simple, in=120, out=0, looseness=1.00] (26.center) to (32);
	\end{pgfonlayer}
\end{tikzpicture}
\hfill\]
This implies in particular that $Z$ and $X$ phase gates commute through spiders of the same colour. Pauli $Z$ and $X$ gates can be pushed through spiders of the opposite colour, but they change the sign of the angle:
\begin{equation}\label{eq:z-stab}
  \hfill\begin{tikzpicture}
	\begin{pgfonlayer}{nodelayer}
		\node [style=white phase dot] (0) at (-2.75, 0) {$\alpha$};
		\node [style=none] (1) at (-1.25, 1.5) {};
		\node [style=gray phase dot] (2) at (-4.5, 0) {$\pi$};
		\node [style=none] (3) at (-1.25, -1.25) {};
		\node [style=none] (4) at (-1.25, 0.5) {};
		\node [style=none, rotate=90] (5) at (-1.5, -0.25) {...};
		\node [style=none] (6) at (0, 0) {$=$};
		\node [style=none] (7) at (-5.5, 0) {};
		\node [style=white phase dot] (8) at (2.75, 0) {-$\alpha$};
		\node [style=gray phase dot] (9) at (4.25, 0.5) {$\pi$};
		\node [style=none] (10) at (1.25, 0) {};
		\node [style=none, rotate=90] (11) at (3.75, -0.25) {...};
		\node [style=none] (12) at (5, 0.5) {};
		\node [style=gray phase dot] (13) at (4.25, 1.5) {$\pi$};
		\node [style=none] (14) at (5, -1.25) {};
		\node [style=gray phase dot] (15) at (4.25, -1.25) {$\pi$};
		\node [style=none] (16) at (5, 1.5) {};
	\end{pgfonlayer}
	\begin{pgfonlayer}{edgelayer}
		\draw [style=simple, in=180, out=-60, looseness=0.75] (0) to (3.center);
		\draw [style=simple, in=180, out=45, looseness=0.75] (0) to (4.center);
		\draw [style=simple, in=180, out=75, looseness=0.75] (0) to (1.center);
		\draw [style=simple, in=180, out=0, looseness=0.50] (2) to (0);
		\draw [style=simple] (7.center) to (2);
		\draw [style=simple, in=-60, out=180, looseness=0.75] (15) to (8);
		\draw [style=simple, in=0, out=180, looseness=0.75] (8) to (10.center);
		\draw [style=simple, in=180, out=45, looseness=0.75] (8) to (9);
		\draw [style=simple, in=75, out=180, looseness=0.75] (13) to (8);
		\draw [style=simple] (16.center) to (13);
		\draw [style=simple] (12.center) to (9);
		\draw [style=simple] (14.center) to (15);
	\end{pgfonlayer}
\end{tikzpicture}\hfill
\end{equation}
\begin{equation}\label{eq:x-stab}
  \hfill\begin{tikzpicture}
	\begin{pgfonlayer}{nodelayer}
		\node [style={gray phase dot}] (0) at (-2.75, 0) {$\alpha$};
		\node [style=none] (1) at (-1.25, 1.5) {};
		\node [style=none] (2) at (-1.25, -1.25) {};
		\node [style=none] (3) at (-1.25, 0.5) {};
		\node [style={white phase dot}] (4) at (-4.25, -0) {$\pi$};
		\node [style=none, rotate=90] (5) at (-1.5, -0.25) {...};
		\node [style=none] (6) at (0, 0) {$=$};
		\node [style=none] (7) at (-5.25, -0) {};
		\node [style={gray phase dot}] (8) at (2.75, 0) {-$\alpha$};
		\node [style={white phase dot}] (9) at (4.25, 0.5) {$\pi$};
		\node [style=none] (10) at (1.25, -0) {};
		\node [style=none, rotate=90] (11) at (3.75, -0.25) {...};
		\node [style=none] (12) at (5.25, 0.5) {};
		\node [style={white phase dot}] (13) at (4.25, 1.5) {$\pi$};
		\node [style=none] (14) at (5.25, -1.25) {};
		\node [style={white phase dot}] (15) at (4.25, -1.25) {$\pi$};
		\node [style=none] (16) at (5.25, 1.5) {};
	\end{pgfonlayer}
	\begin{pgfonlayer}{edgelayer}
		\draw [style=simple, in=180, out=-60, looseness=0.75] (0) to (2.center);
		\draw [style=simple, in=180, out=45, looseness=0.75] (0) to (3.center);
		\draw [style=simple, in=180, out=75, looseness=0.75] (0) to (1.center);
		\draw [style=simple, in=0, out=180, looseness=0.75] (0) to (4);
		\draw [style=simple] (7.center) to (4);
		\draw [style=simple, in=-60, out=180, looseness=0.75] (15) to (8);
		\draw [style=simple, in=0, out=180, looseness=0.75] (8) to (10.center);
		\draw [style=simple, in=180, out=45, looseness=0.75] (8) to (9);
		\draw [style=simple, in=75, out=180, looseness=0.75] (13) to (8);
		\draw [style=simple] (16.center) to (13);
		\draw [style=simple] (12.center) to (9);
		\draw [style=simple] (14.center) to (15);
	\end{pgfonlayer}
\end{tikzpicture}\hfill
\end{equation}

We introduce a special notation for the Hadamard gate:
\begin{equation}\label{eq:had-def}
\hfill \frac{1}{\sqrt{2}}\left(\begin{array}{cc} 1 & 1 \\ 1 & -1 \end{array}\right) \ =\ \ \begin{tikzpicture}
	\begin{pgfonlayer}{nodelayer}
		\node [style=none] (0) at (-0.25, 0) {};
		\node [style=none] (1) at (1.75, 0) {};
		\node [style=small hadamard] (2) at (0.75, 0) {};
		\node [style=none] (3) at (2.5, 0) {$=$};
		\node [style=none] (4) at (3.25, 0) {};
		\node [style=gray phase dot] (5) at (4, 0) {-$\frac\pi2$};
		\node [style=white phase dot] (6) at (5, 0) {-$\frac\pi2$};
		\node [style=gray phase dot] (7) at (6, 0) {-$\frac\pi2$};
		\node [style=none] (8) at (6.75, 0) {};
	\end{pgfonlayer}
	\begin{pgfonlayer}{edgelayer}
		\draw [style=simple] (0.center) to (2);
		\draw [style=simple] (2) to (1.center);
		\draw [style=simple] (4.center) to (8.center);
	\end{pgfonlayer}
\end{tikzpicture}
\end{equation}

And of course the Hadamard is self-inverse:
\[\hfill \begin{tikzpicture}
	\begin{pgfonlayer}{nodelayer}
		\node [style=none] (0) at (-0.5, 0) {};
		\node [style=none] (1) at (1.75, 0) {};
		\node [style=small hadamard] (2) at (1, 0) {};
		\node [style=none] (3) at (2.5, 0) {$=$};
		\node [style=small hadamard] (4) at (0.25, 0) {};
		\node [style=none] (5) at (5.5, 0) {};
		\node [style=none] (6) at (3.25, 0) {};
	\end{pgfonlayer}
	\begin{pgfonlayer}{edgelayer}
		\draw (0.center) to (2);
		\draw (2) to (1.center);
		\draw (6.center) to (5.center);
	\end{pgfonlayer}
\end{tikzpicture} \hfill\]
The Hadamard gate interchanges the $Z$ and $X$ bases, so that it acts as a colour-changer for spiders:
\ctikzfig{colour-ch2}

A final property that we will use is the \emph{strong complementarity} of the Z and X bases of a qubit, which are captured by the following two diagrammatic rules:
\ctikzfig{strong-compl}
We will refer to these two rules, respectively, as the \textit{exchange} rule and the \textit{copy} rule.

Combining the rightmost equation with the other rules yields a more general version, for any $a \in \{0,1\}$ and $\alpha \in [0,2\pi)$:
\ctikzfig{gen-copy}

\begin{figure}
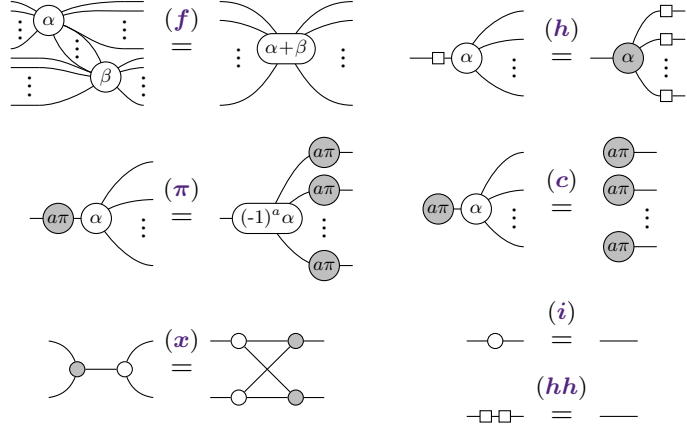

\ctikzfig{ZX-rules}
\caption{The rules of the ZX-calculus: spider-\SpiderRule{}usion, \HadamardRule{}adamard, \PiRule{}-commutation, \CopyRule{}opy, e\HopfRule{}change, \IdRule{}dentity, and \HHRule{}-cancellation. Note these rules hold for all $\alpha, \beta \in [0, 2 \pi)$ and $a \in \{0,1\}$ and also with the colours interchanged, due to \HadamardRule and \HHRule.}\label{table:rewrite}
\end{figure}

We summarise all these rules in Figure~\ref{table:rewrite}. Note that an extended set of these rules has been proven to be \emph{complete} \cite{HarnyCompleteness,SimonCompleteness}, that is, two diagrams represent the same linear map if and only if they can be rewritten into one another using the rules of the ZX-calculus. In contrast, the rules of Figure~\ref{table:rewrite} are only complete when restricted to the Clifford fragment~\cite{Backens1}.

As a demonstration of rewriting with these rules, let us derive the following equality, which we will use throughout the paper:
\begin{equation}\label{eq:s-state-eq}
  \hfill\begin{tikzpicture}
	\begin{pgfonlayer}{nodelayer}
		\node [style=white phase dot] (0) at (-10, 0) {$\frac\pi2$};
		\node [style=none] (1) at (-8.75, 0) {};
		\node [style=none] (2) at (-8, 0) {=};
		\node [style=none] (3) at (-4.75, 0) {};
		\node [style=small hadamard] (4) at (-5.75, 0) {};
		\node [style=gray phase dot] (5) at (-6.75, 0) {$\frac\pi2$};
		\node [style=none] (6) at (1.5, 0) {};
		\node [style=gray phase dot] (7) at (-3, 0) {$\frac\pi2$};
		\node [style=none] (8) at (-4, 0) {=};
		\node [style=gray phase dot] (9) at (-1.75, 0) {-$\frac\pi2$};
		\node [style=white phase dot] (10) at (-0.5, 0) {-$\frac\pi2$};
		\node [style=gray phase dot] (11) at (0.75, 0) {-$\frac\pi2$};
		\node [style=none] (12) at (2.25, 0) {=};
		\node [style=gray dot] (13) at (3, 0) {};
		\node [style=gray phase dot] (14) at (5.25, 0) {-$\frac\pi2$};
		\node [style=white phase dot] (15) at (4, 0) {-$\frac\pi2$};
		\node [style=none] (16) at (6, 0) {};
		\node [style=none] (17) at (6.75, 0) {$=$};
		\node [style=gray dot] (18) at (7.5, 0) {};
		\node [style=gray phase dot] (19) at (8.25, 0) {-$\frac\pi2$};
		\node [style=none] (20) at (9, 0) {};
		\node [style=none] (21) at (9.75, 0) {=};
		\node [style=gray phase dot] (22) at (10.75, 0) {-$\frac\pi2$};
		\node [style=none] (23) at (11.75, 0) {};
		\node [style=none] (24) at (-8, 1) {\hadamardrule};
		\node [style=none] (25) at (-4, 1) {\eqref{eq:had-def}};
		\node [style=none] (26) at (2.25, 1) {\spiderrule};
		\node [style=none] (27) at (6.75, 1) {\pirule};
		\node [style=none] (28) at (9.75, 1) {\spiderrule};
	\end{pgfonlayer}
	\begin{pgfonlayer}{edgelayer}
		\draw [style=simple] (0) to (1.center);
		\draw [style=simple] (5) to (4);
		\draw [style=simple] (4) to (3.center);
		\draw [style=simple] (7) to (9);
		\draw [style=simple] (9) to (10);
		\draw [style=simple] (10) to (11);
		\draw [style=simple] (11) to (6.center);
		\draw [style=simple] (15) to (14);
		\draw [style=simple] (14) to (16.center);
		\draw [style=simple] (13) to (15);
		\draw [style=simple] (19) to (20.center);
		\draw [style=simple] (18) to (19);
		\draw [style=simple] (22) to (23.center);
	\end{pgfonlayer}
\end{tikzpicture}\hfill
\end{equation}
The same equation with the colours reversed also holds.

\setlength{\tabcolsep}{10pt}

\section{PPM in the ZX-calculus}\label{sec:ppm-zx}

In order to express the PPM model in the ZX-calculus, we give graphical presentations for the parity-phase gate and for Pauli measurements.
Recall that $P(\alpha) = \exp(-i\frac\alpha2 Z\otimes Z)$. It is straightforward to check that as a matrix in the computational basis, we have:
\[\hfill \exp(-i\frac\alpha2 Z\otimes Z) =  e^{-i\frac\alpha2}\left(\begin{matrix}
  1  & 0 & 0 & 0 \\
  0 & e^{i \alpha} & 0 & 0 \\
  0 & 0 & e^{i \alpha} & 0 \\
  0 & 0 & 0 & 1  \\
\end{matrix}\right)\hfill\]
If we interchange the third and fourth computational state ($\ket{10}$ and $\ket{11}$), we obtain (up to a global phase):
\[\hfill I\otimes Z(\alpha) = \left(\begin{matrix}
  1  & 0 & 0 & 0 \\
  0 & e^{i \alpha} & 0 & 0 \\
  0 & 0 & 1 & 0 \\
  0 & 0 & 0 & e^{i \alpha}  \\
\end{matrix}\right)
\hfill\]
Hence we obtain the parity-phase gate by pre- and post-composing the above matrix with CNOTs:
\[\hfill P(\alpha) = \text{CNOT}(I\otimes Z(\alpha))\text{CNOT} \ =\ \begin{tikzpicture}
	\begin{pgfonlayer}{nodelayer}
		\node [style=white dot] (0) at (-1, 0.5) {};
		\node [style=none] (1) at (2.25, 0.5) {};
		\node [style=none] (2) at (-2.25, -0.5) {};
		\node [style=gray dot] (3) at (-1, -0.5) {};
		\node [style=none] (4) at (2.25, -0.5) {};
		\node [style=white phase dot] (5) at (0, -0.5) {$\alpha$};
		\node [style=gray dot] (6) at (1, -0.5) {};
		\node [style=none] (7) at (-2.25, 0.5) {};
		\node [style=white dot] (8) at (1, 0.5) {};
	\end{pgfonlayer}
	\begin{pgfonlayer}{edgelayer}
		\draw [style=simple] (0) to (3);
		\draw [style=simple] (0) to (7.center);
		\draw [style=simple] (2.center) to (3);
		\draw [style=simple] (8) to (6);
		\draw [style=simple] (8) to (1.center);
		\draw [style=simple] (6) to (4.center);
		\draw (0) to (8);
		\draw (3) to (5);
		\draw (5) to (6);
	\end{pgfonlayer}
\end{tikzpicture} \hfill\]
This simplifies as a routine calculation in the ZX-calculus (see e.g.~Ref.~\cite{CKbook}):
\ctikzfig{MS-simplify}
Hence, we have:
\begin{equation}\label{eq:virtual-qubit}
  \hfill P(\alpha) \ =\ \begin{tikzpicture}
	\begin{pgfonlayer}{nodelayer}
		\node [style=none] (0) at (-8.25, -1) {};
		\node [style=none] (1) at (-5.5, -1) {};
		\node [style=none] (2) at (-8.25, 1) {};
		\node [style=none] (3) at (-5.5, 1) {};
		\node [style=white dot] (4) at (-7, 1) {};
		\node [style=white dot] (5) at (-7, -1) {};
		\node [style=gray dot] (6) at (-7, 0) {};
		\node [style=white phase dot] (7) at (-8, 0) {$\alpha$};
	\end{pgfonlayer}
	\begin{pgfonlayer}{edgelayer}
		\draw [style=simple] (0.center) to (5.center);
		\draw [style=simple] (5.center) to (1.center);
		\draw [style=simple] (5.center) to (6.center);
		\draw [style=simple] (6.center) to (4.center);
		\draw [style=simple] (4.center) to (2.center);
		\draw [style=simple] (4.center) to (3.center);
		\draw [style=simple] (7.center) to (6.center);
	\end{pgfonlayer}
\end{tikzpicture}\hfill
\end{equation}

The Pauli Z and X measurements non-deterministically introduce projections onto their respective eigenstates, namely $\{ \bra{0}, \bra{1} \}$ for $Z$-measurements and $\{ \bra{+}, \bra{-} \}$ for $X$-measurements. Because basis elements of one colour can be expressed as spiders of the other colour, we can depict measurements as follows:
\begin{equation}\label{eq:meas-effects}
\begin{split}
  \textit{\small Z-measure} & := \left\{ \begin{tikzpicture}
	\begin{pgfonlayer}{nodelayer}
		\node [style=none] (0) at (-0.5, 0) {};
		\node [style=gray phase dot] (1) at (0.25, 0) {$a \pi$};
	\end{pgfonlayer}
	\begin{pgfonlayer}{edgelayer}
		\draw [style=simple] (0.center) to (1);
	\end{pgfonlayer}
\end{tikzpicture}\right\}_{a\in\{0,1\}} \\
  \textit{\small X-measure} & := \left\{\begin{tikzpicture}
	\begin{pgfonlayer}{nodelayer}
		\node [style=none] (0) at (-0.5, 0) {};
		\node [style=white phase dot] (1) at (0.25, 0) {$a \pi$};
	\end{pgfonlayer}
	\begin{pgfonlayer}{edgelayer}
		\draw [style=simple] (0.center) to (1);
	\end{pgfonlayer}
\end{tikzpicture}\right\}_{a\in\{0,1\}}
\end{split}
\end{equation}

The appearance of the `virtual qubit', i.e. the state being input between the two wires in equation \eqref{eq:virtual-qubit} is quite suggestive. We can interpret this node as a magic state that is waiting to be applied to one of its neighbouring (actual) qubits to introduce a phase. More specifically, if we prepare the second qubit in the $\ket{+}$ state and then measure it in the Z basis, we obtain a Z-phase gate $Z(\pm \alpha)$, with the sign depending on the measurement outcome:
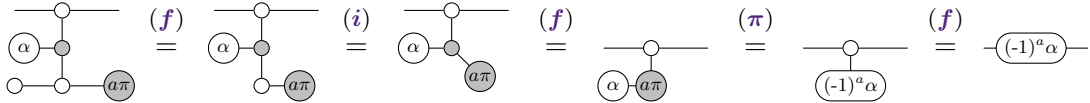
\ctikzfig{zx-msgate-magic}

If we take $\alpha=\frac\pi2$ we can rewrite the expression of $P(\frac\pi2)$ in the ZX-calculus a bit further:
\begin{equation}\label{eq:mscx}
  \hfill\begin{tikzpicture}
	\begin{pgfonlayer}{nodelayer}
		\node [style=gray phase dot] (0) at (-3.75, 0) {-$\frac\pi2$};
		\node [style=none] (1) at (-6.75, -1) {};
		\node [style=none] (2) at (-9.5, -1) {};
		\node [style=none] (3) at (-9.5, 1) {};
		\node [style=none] (4) at (-6.75, 1) {};
		\node [style=none] (5) at (-5.5, 0) {=};
		\node [style=white phase dot] (6) at (-9, 0) {$\frac\pi2$};
		\node [style=none] (7) at (-0.5, 0) {=};
		\node [style=none] (8) at (3, -2) {};
		\node [style=none] (9) at (3, 2) {};
		\node [style=none] (10) at (0.25, -2) {};
		\node [style=none] (11) at (0.25, 2) {};
		\node [style=gray phase dot] (12) at (1.5, 0) {-$\frac\pi2$};
		\node [style=none] (13) at (4, 0) {=};
		\node [style=none] (14) at (8, 1) {};
		\node [style=none] (15) at (8, -1) {};
		\node [style=none] (16) at (5.25, 1) {};
		\node [style=none] (17) at (5.25, -1) {};
		\node [style=small hadamard] (18) at (6.5, 0) {};
		\node [style=none] (19) at (-10.75, 0) {=};
		\node [style=none] (20) at (-13, 0) {$P(\frac\pi2)$};
		\node [style=white dot] (21) at (-8, 1) {};
		\node [style=white dot] (22) at (-8, -1) {};
		\node [style=gray dot] (23) at (-8, 0) {};
		\node [style=none] (24) at (-4.25, -1) {};
		\node [style=none] (25) at (-1.5, 1) {};
		\node [style=white dot] (26) at (-2.75, 1) {};
		\node [style=white dot] (27) at (-2.75, -1) {};
		\node [style=gray dot] (28) at (-2.75, 0) {};
		\node [style=none] (29) at (-4.25, 1) {};
		\node [style=none] (30) at (-1.5, -1) {};
		\node [style=white phase dot] (31) at (1.5, 1) {-$\frac\pi2$};
		\node [style=white phase dot] (32) at (1.5, 2) {$\frac\pi2$};
		\node [style=white phase dot] (33) at (1.5, -1) {-$\frac\pi2$};
		\node [style=white phase dot] (34) at (1.5, -2) {$\frac\pi2$};
		\node [style=white phase dot] (35) at (6.5, 1) {$\frac\pi2$};
		\node [style=white phase dot] (36) at (6.5, -1) {$\frac\pi2$};
		\node [style=none] (37) at (-5.5, 0.75) {\eqref{eq:s-state-eq}};
		\node [style=none] (38) at (-0.5, 0.75) {\spiderrule};
		\node [style=none] (39) at (4, 0.75) {\eqref{eq:had-def}};
	\end{pgfonlayer}
	\begin{pgfonlayer}{edgelayer}
		\draw [style=simple] (1.center) to (22);
		\draw [style=simple] (22) to (2.center);
		\draw [style=simple] (22) to (23);
		\draw [style=simple] (23) to (21);
		\draw [style=simple] (21) to (4.center);
		\draw [style=simple] (21) to (3.center);
		\draw [style=simple] (30.center) to (27);
		\draw [style=simple] (27) to (24.center);
		\draw [style=simple] (27) to (28);
		\draw [style=simple] (28) to (26);
		\draw [style=simple] (26) to (25.center);
		\draw [style=simple] (26) to (29.center);
		\draw [style=simple] (23) to (6);
		\draw [style=simple] (28) to (0);
		\draw [style=simple] (11.center) to (32);
		\draw [style=simple] (32) to (9.center);
		\draw [style=simple] (32) to (31);
		\draw [style=simple] (31) to (12);
		\draw [style=simple] (12) to (33);
		\draw [style=simple] (33) to (34);
		\draw [style=simple] (34) to (10.center);
		\draw [style=simple] (34) to (8.center);
		\draw [style=simple] (17.center) to (36);
		\draw [style=simple] (36) to (15.center);
		\draw [style=simple] (36) to (18);
		\draw [style=simple] (18) to (35);
		\draw [style=simple] (35) to (16.center);
		\draw [style=simple] (35) to (14.center);
	\end{pgfonlayer}
\end{tikzpicture} \hfill
\end{equation}
Written in this way it is clear that the gate is equivalent, up to single-qubit Clifford unitaries to the controlled-Z gate $CZ$, which is represented in ZX-notation as:
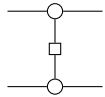
\[\hfill\begin{tikzpicture}
	\begin{pgfonlayer}{nodelayer}
		\node [style=none] (0) at (1.25, 1) {};
		\node [style=none] (1) at (1.25, -1) {};
		\node [style=none] (2) at (-1.25, 1) {};
		\node [style=none] (3) at (-1.25, -1) {};
		\node [style=small hadamard] (4) at (0, 0) {};
		\node [style=white dot] (5) at (0, 1) {};
		\node [style=white dot] (6) at (0, -1) {};
	\end{pgfonlayer}
	\begin{pgfonlayer}{edgelayer}
		\draw [style=simple] (3.center) to (6);
		\draw [style=simple] (6) to (1.center);
		\draw [style=simple] (6) to (4);
		\draw [style=simple] (4) to (5);
		\draw [style=simple] (5) to (2.center);
		\draw [style=simple] (5) to (0.center);
	\end{pgfonlayer}
\end{tikzpicture} \hfill\]

Since $P(\alpha)P(\beta) = P(\alpha+\beta)$ we get $P(\frac\pi2) = P(\frac\pi4)^2$, so that the gate $P(\frac\pi4)$ is a $\sqrt{CZ}$ gate, up to local unitaries.

\begin{convention*}
  From hence forth, we will typically suppress explicit references to the spider-fusion rule \SpiderRule, and assume that spiders of the same colour are (un)fused as necessary. Similarly, we will suppress references to \IdRule and \HHRule and simply remove 2-legged spiders and pairs of $H$ gates as they appear, as both are equal to the identity matrix.
\end{convention*}


We can now define the translation from P-graph states and patterns to ZX-diagrams, which is similar in spirit to the one given in Ref.~\cite{DP2} for graph states.
Qubits become white dots with a single output and single/double edges become edges decorated by the appropriate phases as follows:
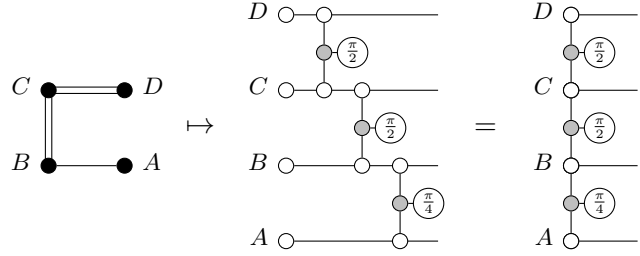
\ctikzfig{ms-translate}

It will be convenient to deform the right-hand side to match the topology of the associated P-graph state, in which case we can drop the qubit labels:
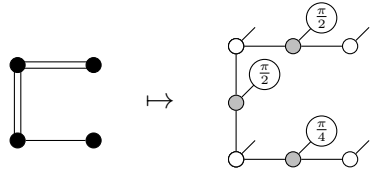
\ctikzfig{ms-translate2}
Note that all of the wires with a free end correspond to outputs, so there is no need to draw them exiting to the right of the diagram.

To compute the result of a measurement pattern, we post-compose with the appropriate effects $\{ \bra{0}, \bra{1} \}$ for $Z$-measurements and $\{ \bra{+}, \bra{-} \}$ for $X$-measurements using equation \eqref{eq:meas-effects}. This enables us to write patterns (without feed-forward) as a single ZX-diagram:
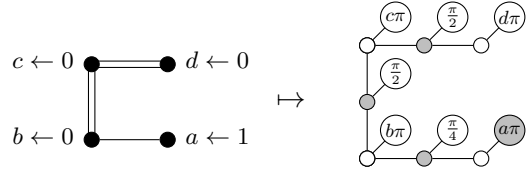
\ctikzfig{ms-translate3}
We could also represent the feed-forward within the diagram (e.g. by conditionally applying Hadamard gates to outputs), but for our purposes, it will be simpler just to do some simple case-distinctions.

Finally, pattern fragments can be expressed by not measuring outputs, and adding a new input wire for each input:
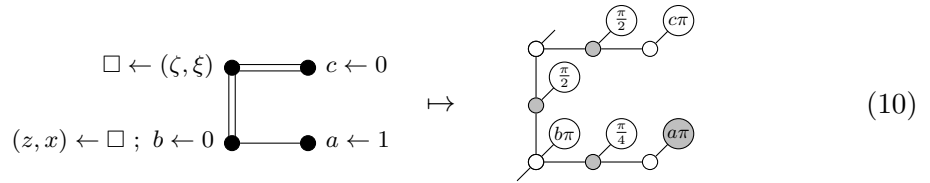
\begin{equation}\label{eq:pattern-translate}
  \hfill\begin{tikzpicture}
	\begin{pgfonlayer}{nodelayer}
		\node [style=small black dot] (0) at (-5.5, -1) {};
		\node [style=small black dot] (1) at (-3.5, 1) {};
		\node [style=small black dot] (2) at (-5.5, 1) {};
		\node [style=small black dot] (3) at (-3.5, -1) {};
		\node [style=none] (4) at (0, 0) {$\mapsto$};
		\node [style=white dot] (5) at (5.5, 1.5) {};
		\node [style=white dot] (6) at (2.5, 1.5) {};
		\node [style=white phase dot] (7) at (4.75, -0.75) {$\frac\pi4$};
		\node [style=white phase dot] (8) at (3.25, 0.75) {$\frac\pi2$};
		\node [style=white dot] (9) at (2.5, -1.5) {};
		\node [style=white dot] (10) at (5.5, -1.5) {};
		\node [style=white dot] (11) at (2.5, -1.5) {};
		\node [style=white dot] (12) at (2.5, -1.5) {};
		\node [style=white phase dot] (13) at (6.25, 2.25) {$c\pi$};
		\node [style=white dot] (14) at (2.5, 1.5) {};
		\node [style=gray dot] (15) at (4, -1.5) {};
		\node [style=gray dot] (16) at (2.5, 0) {};
		\node [style=white phase dot] (17) at (3.25, -0.75) {$b\pi$};
		\node [style=none] (18) at (3, 2) {};
		\node [style=gray phase dot] (19) at (6.25, -0.75) {$a\pi$};
		\node [style=gray dot] (20) at (4, 1.5) {};
		\node [style=white phase dot] (21) at (4.75, 2.25) {$\frac\pi2$};
		\node [style=white dot] (22) at (2.5, 1.5) {};
		\node [style=left label] (23) at (-6, 1) {$\square \leftarrow (\zeta, \xi)$};
		\node [style=right label] (24) at (-3, 1) {$c \leftarrow 0$};
		\node [style=left label] (25) at (-6, -1) {$(z,x) \leftarrow \square\ ;\ b\leftarrow 0$};
		\node [style=right label] (26) at (-3, -1) {$a \leftarrow 1$};
		\node [style=none] (27) at (2, -2) {};
	\end{pgfonlayer}
	\begin{pgfonlayer}{edgelayer}
		\draw [style=gs double edge] (2) to (1);
		\draw [style=gs double edge] (0) to (2);
		\draw [style=gs edge] (0) to (3);
		\draw (5) to (13);
		\draw (6) to (18.center);
		\draw (11) to (17);
		\draw (10) to (19);
		\draw (20) to (14);
		\draw (20) to (21);
		\draw (22) to (16);
		\draw (16) to (8);
		\draw (16) to (9);
		\draw (12) to (15);
		\draw (15) to (7);
		\draw (20) to (5);
		\draw (15) to (10);
		\draw (27.center) to (12);
	\end{pgfonlayer}
\end{tikzpicture}\hfill
\end{equation}

In order to implement a gate $G$, we should show that the right-hand side above, pre-composed with possible Pauli errors, implements $G$ followed by some possible Pauli errors. We can represent the possible Pauli errors as follows:
\begin{equation}\label{eq:errors}
  \hfill\begin{tikzpicture}
	\begin{pgfonlayer}{nodelayer}
		\node [style=none] (0) at (-2.75, 0) {};
		\node [style=white phase dot] (1) at (-3.5, 0) {$z \pi$};
		\node [style=left label, anchor=east] (2) at (4.75, 0) {$\textit{$X$-error} := \bigg\{$};
		\node [style=left label, anchor=west] (3) at (6.25, 0) {$\bigg\}_{x\in \{0,1\}}$};
		\node [style=gray phase dot] (4) at (5.75, 0) {$x \pi$};
		\node [style=left label, anchor=west] (5) at (-3, 0) {$\bigg\}_{z\in \{0,1\}}$};
		\node [style=none] (6) at (6.5, 0) {};
		\node [style=left label, anchor=east] (7) at (-4.5, 0) {$\textit{$Z$-error} := \bigg\{$};
		\node [style=none] (8) at (-4.25, 0) {};
		\node [style=none] (9) at (5, 0) {};
	\end{pgfonlayer}
	\begin{pgfonlayer}{edgelayer}
		\draw [style=simple] (0.center) to (1);
		\draw [style=simple] (6.center) to (4);
		\draw [style=simple, in=360, out=180] (1) to (8.center);
		\draw [style=simple, in=360, out=180] (4) to (9.center);
	\end{pgfonlayer}
\end{tikzpicture}\hfill
\end{equation}
Then, giving a deterministic implementation of a gate $G$ amounts to proving that there exists boolean functions $\zeta, \xi$ such that the following equation holds, for all values of the boolean variables $a, b, c, x, z$:
\ctikzfig{ms-translate5}

\begin{remark*}
  The colours play opposite roles in equations \eqref{eq:meas-effects} and \eqref{eq:errors}. This is a common theme in the ZX-calculus, and comes from the fact that basis elements of one colour can be described as spiders of the other colour (cf. equation~\eqref{eq:spider-basis1}).
\end{remark*}

\section{Measurement patterns for a universal set of gates}

In this section, we will introduce several pattern fragments, and show using the ZX-calculus that they deterministically implement certain quantum gates. We will start a simple example, which uses one double-edge to implement an $X(\pi/2) = HSH$ gate. Following that, we will show that a single P-graph, with different measurement patterns, can implement $T$, $H$, or $S$ gates. Similarly, we show a different P-graph shape can be used to selectively implement CZ or $S \otimes S$. These will be used in the next section to show universality of the PPM model.

The pattern for an $HSH$ gate is:
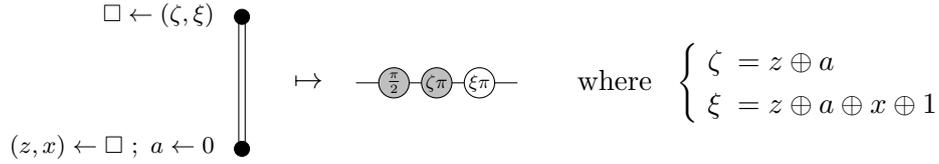
\begin{equation*}\label{eq:pat-HSH}
\begin{tikzpicture}
	\begin{pgfonlayer}{nodelayer}
		\node [style=small black dot] (0) at (-1.75, -1.75) {};
		\node [style=small black dot] (1) at (-1.75, 1.75) {};
		\node [style=none] (2) at (0, 0) {$\mapsto$};
		\node [style=gray phase dot] (3) at (2.25, 0) {$\frac\pi2$};
		\node [style=none] (4) at (5.5, 0) {};
		\node [style=none] (5) at (1.25, 0) {};
		\node [style=gray phase dot] (6) at (3.375, 0) {$\zeta \pi$};
		\node [style=white phase dot] (7) at (4.5, 0) {$\xi \pi$};
		\node [style=right label] (8) at (-1, 1.75) {};
		\node [style=left label] (9) at (-2.5, -1.75) {$(z,x) \leftarrow \square\ ;\ a\leftarrow 0$};
		\node [style=left label] (10) at (-2.5, 1.75) {$\square \leftarrow (\zeta,\xi)$};
	\end{pgfonlayer}
	\begin{pgfonlayer}{edgelayer}
		\draw [style=gs double edge] (0) to (1);
		\draw [style=simple] (5.center) to (3);
		\draw [style=simple] (3) to (4.center);
	\end{pgfonlayer}
\end{tikzpicture}
\qquad\textrm{where}\ \ 
\begin{cases}
  \ \zeta & \!\!\!=  z\oplus a\\
  \ \xi & \!\!\!= z\oplus a\oplus x\oplus 1
\end{cases}
\end{equation*}

The bottom qubit is the input of the expression. We always measure it in the X basis ($a\leftarrow 0$) which gives us a measurement result $a$. We record the incoming $Z$ and $X$ error in the variables $z$ and $x$. The resulting $Z$ error at the end is now $z\oplus a$, and the X error is $z\oplus a\oplus x\oplus 1$. We can show the correctness of this fragment by performing translation \eqref{eq:pattern-translate} and reducing using the ZX-calculus:
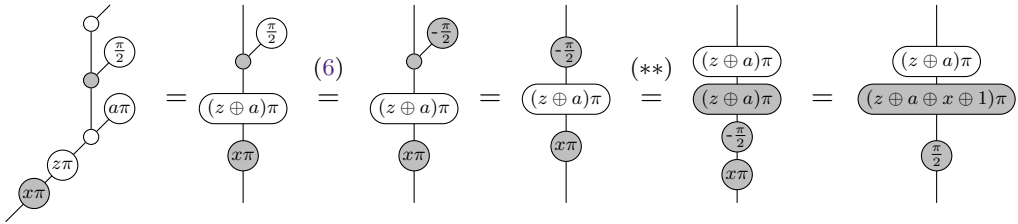
\ctikzfig{ms-sgate-zx}
Here equation $(**)$ is the standard Clifford commutation law of $S^\dagger X \propto X Z S^\dagger $. This commutation follows straightforwardly from \PiRule and the fact that, for $a \in \{0,1\}$, we have $(-1)^a \frac\pi2 = \frac\pi2 + a\pi$ (mod $2\pi$).

\begin{equation}\label{eq:S-commute}
    \hfill\begin{tikzpicture}
	\begin{pgfonlayer}{nodelayer}
		\node [style=none] (0) at (-3.5, -2) {};
		\node [style=none] (1) at (-3.5, 2.25) {};
		\node [style=white phase dot] (2) at (-3.5, -0.75) {$\frac\pi2$};
		\node [style=gray phase dot] (3) at (-3.5, 0.75) {$a\pi$};
		\node [style=none] (4) at (0, -2) {};
		\node [style=none] (5) at (0, 2.25) {};
		\node [style=white phase dot] (6) at (0, 0.75) {$\ (\textrm{-}1)^a\frac\pi2\ $};
		\node [style=gray phase dot] (7) at (0, -0.75) {$a\pi$};
		\node [style=none] (8) at (-2, 0) {=};
		\node [style=none] (9) at (-2, 0.75) {\pirule};
		\node [style=none] (10) at (7.25, -2) {};
		\node [style=none] (11) at (7.25, 2.25) {};
		\node [style=white phase dot] (12) at (7.25, 1.5) {$\frac\pi2$};
		\node [style=gray phase dot] (13) at (7.25, -0.75) {$a\pi$};
		\node [style=none] (14) at (5.75, 0) {=};
		\node [style=white phase dot] (16) at (7.25, 0.5) {$a\pi$};
		\node [style=none] (17) at (4, -2) {};
		\node [style=none] (18) at (4, 2.25) {};
		\node [style=white phase dot] (19) at (4, 0.75) {$\ \frac\pi2 + a\pi\ $};
		\node [style=gray phase dot] (20) at (4, -0.75) {$a\pi$};
		\node [style=none] (21) at (2, 0) {=};
	\end{pgfonlayer}
	\begin{pgfonlayer}{edgelayer}
		\draw (0.center) to (2);
		\draw (2) to (3);
		\draw (3) to (1.center);
		\draw (4.center) to (6);
		\draw (6) to (7);
		\draw (7) to (5.center);
		\draw (10.center) to (12);
		\draw (12) to (13);
		\draw (13) to (11.center);
		\draw (17.center) to (19);
		\draw (19) to (20);
		\draw (20) to (18.center);
	\end{pgfonlayer}
\end{tikzpicture}\hfill
\end{equation}


We will now introduce a more versatile P-graph fragment, shaped like an `E', which can implement a variety of single-qubit gates. The first pattern fragment in the E-shape implements a $Z(\pi/4)$ gate, i.e. a $T$ gate:
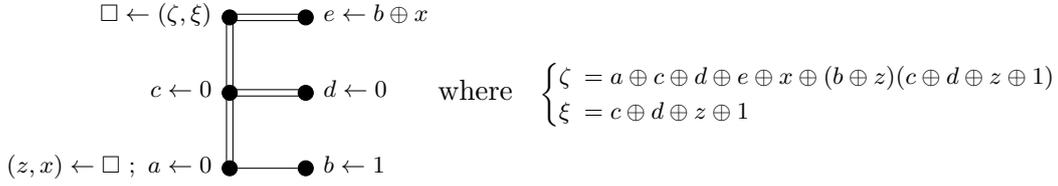
\begin{equation*}\label{eq:pat-T}
\begin{tikzpicture}
	\begin{pgfonlayer}{nodelayer}
		\node [style=mbqc dot] (0) at (0, -2) {};
		\node [style=mbqc dot] (1) at (-2, -2) {};
		\node [style=mbqc dot] (2) at (-2, 0) {};
		\node [style=mbqc dot] (3) at (0, 2) {};
		\node [style=mbqc dot] (4) at (-2, 2) {};
		\node [style=left label] (5) at (-2.5, -2) {$(z,x) \leftarrow \square\ ;\ a\leftarrow 0$};
		\node [style=left label] (6) at (-2.5, 2) {$\square \leftarrow (\zeta, \xi)$};
		\node [style=right label] (7) at (0.5, 2) {$e \leftarrow b\oplus x$};
		\node [style=right label] (8) at (0.5, -2) {$b \leftarrow 1$};
		\node [style=right label] (9) at (0.5, 0) {$d \leftarrow 0$};
		\node [style=mbqc dot] (10) at (0, 0) {};
		\node [style=left label] (11) at (-2.5, 0) {$c \leftarrow 0$};
	\end{pgfonlayer}
	\begin{pgfonlayer}{edgelayer}
		\draw [style=gs double edge] (2) to (1);
		\draw [style=gs edge] (1) to (0);
		\draw [style=gs double edge] (4) to (3);
		\draw [style=gs double edge] (2) to (4);
		\draw [style=gs double edge] (2) to (10);
	\end{pgfonlayer}
\end{tikzpicture}
\textrm{where \footnotesize $\ \ 
\begin{cases}
  \zeta
  & \!\!\! = a \oplus c \oplus d \oplus e \oplus x\oplus (b \oplus z)(c \oplus d \oplus z \oplus 1) \\
  \xi
  & \!\!\! = c \oplus d\oplus z \oplus 1
\end{cases}$}
\end{equation*}

Note that now the basis in which the qubit $e$ is measured depends on the incoming $X$ error and one of the measurement results in the pattern fragment itself. Let's translate this to the ZX-calculus, ignoring for the moment being the Pauli errors that could be at the beginning of the fragment:
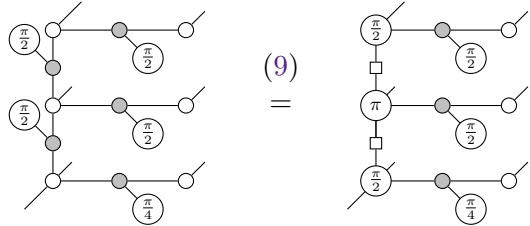
\ctikzfig{ms-zx-tgate1}
Now we measure $a$, $c$ and $d$ in the $X$-basis and $b$ in the $Z$-basis, each of which can introduce a $\pi$ phase of their respective colour:
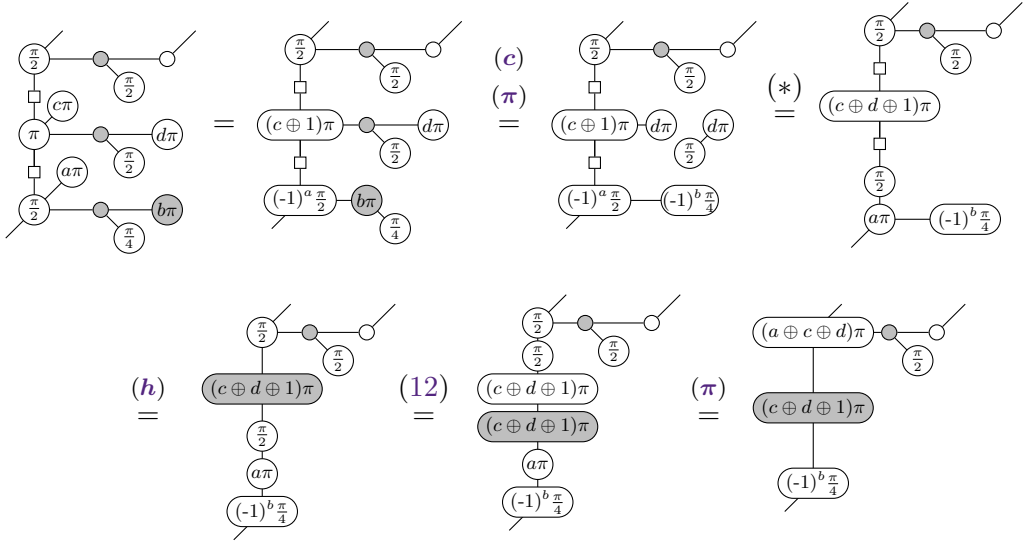
\ctikzfig{ms-zx-tgate2}
In the step marked by ($*$) we have removed the dangling two spider diagram. We are allowed to do this because we are ignoring scalar factors.

The goal of this pattern is to introduce a $\pi/4$ phase. We see that now we either have $\pi/4$ or $-\pi/4$. If we measure $e$ in the X-basis, it gets cut off the main structure, but if we measure it in the Z-basis it introduces an extra $\pi/2$ phase. So, if we got $\pi/4$ (which is the case in the previous diagram when $b=0$), we measure $e$ in the X-basis:
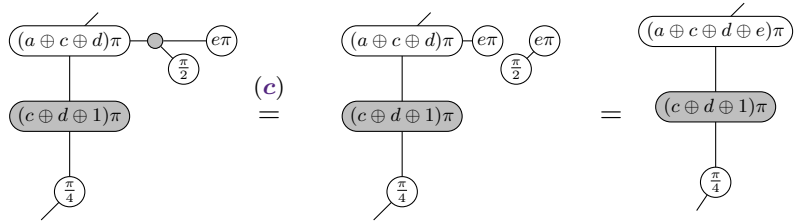
\ctikzfig{ms-zx-tgate3}
and otherwise we measure $e$ in the Z-basis:
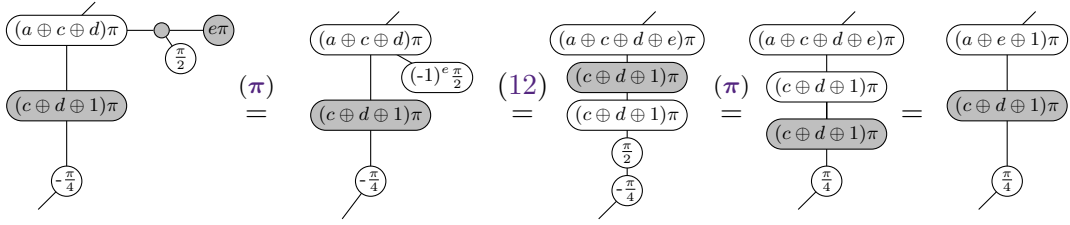
\ctikzfig{ms-zx-tgate4}
so that we indeed implement a $\pi/4$ Z-rotation with some Pauli X and Z error depending on the measurement outcomes. In the presence of of some starting Pauli X and Z errors, the same procedure can be done by being careful to track where the errors spread. Tracking these errors correctly gives the pattern fragment that we started out with. If we decide to measure $e$ in the opposite basis (so in the Z basis when $b=0$ and the X basis when $b=1$), we can implement a $T^\dagger$ gate. Classical control determines the sign of our rotations.

Using the same pattern fragment, but with a different set of measurements on the `hairs' of the fragment we can implement some different operators. For instance, the following pattern fragment gives a Hadamard gate:
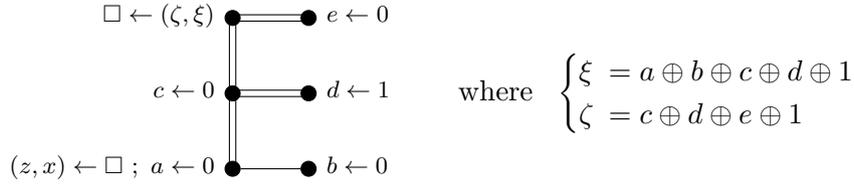
\begin{equation*}\label{eq:pattern-had}
\begin{tikzpicture}
	\begin{pgfonlayer}{nodelayer}
		\node [style=mbqc dot] (0) at (0, -2) {};
		\node [style=mbqc dot] (1) at (-2, -2) {};
		\node [style=mbqc dot] (2) at (-2, 0) {};
		\node [style=mbqc dot] (3) at (0, 2) {};
		\node [style=mbqc dot] (4) at (-2, 2) {};
		\node [style=left label] (5) at (-2.5, -2) {$(z,x) \leftarrow \square\ ;\ a\leftarrow 0$};
		\node [style=left label] (6) at (-2.5, 2) {$\square \leftarrow (\zeta, \xi)$};
		\node [style=right label] (7) at (0.5, 2) {$e \leftarrow 0$};
		\node [style=right label] (8) at (0.5, -2) {$b \leftarrow 0$};
		\node [style=right label] (9) at (0.5, 0) {$d \leftarrow 1$};
		\node [style=mbqc dot] (10) at (0, 0) {};
		\node [style=left label] (11) at (-2.5, 0) {$c \leftarrow 0$};
	\end{pgfonlayer}
	\begin{pgfonlayer}{edgelayer}
		\draw [style=gs double edge] (2) to (1);
		\draw [style=gs edge] (1) to (0);
		\draw [style=gs double edge] (4) to (3);
		\draw [style=gs double edge] (2) to (4);
		\draw [style=gs double edge] (2) to (10);
	\end{pgfonlayer}
\end{tikzpicture}
\qquad\textrm{where} \ \ 
\begin{cases}
\xi & \!\!\!= a \oplus b \oplus c \oplus d \oplus 1 \\
\zeta & \!\!\!= c \oplus d \oplus e \oplus 1
\end{cases}
\end{equation*}
Note there is no feed-forward, so we can verify this in a single derivation:
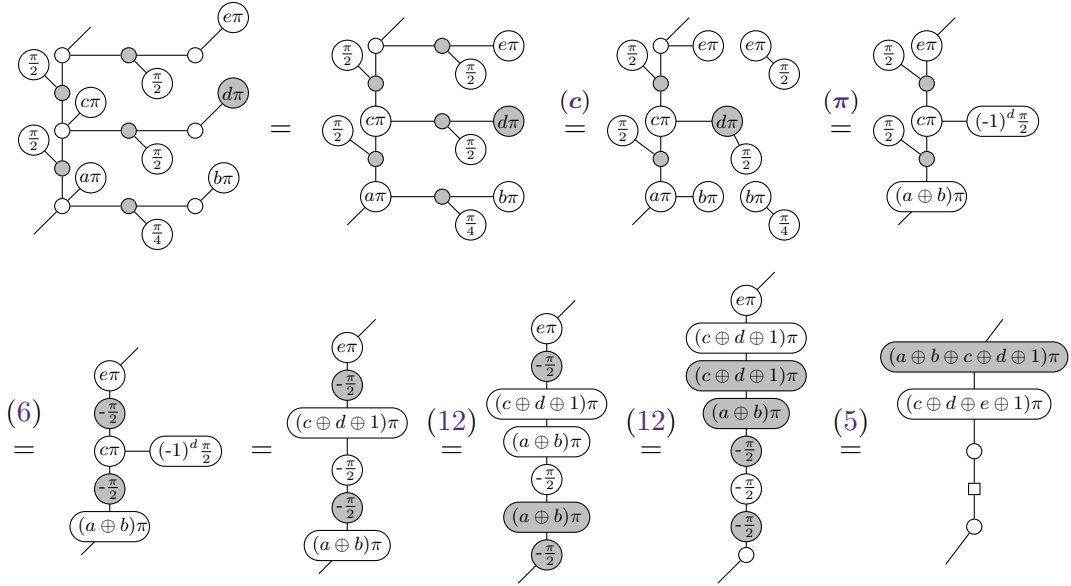
\ctikzfig{ms-zx-had}
In a similar way we can also produce an $S$-gate by measuring qubit $e$ in the Z basis and the rest in the X basis.

The following fragment implements a CZ-gate:
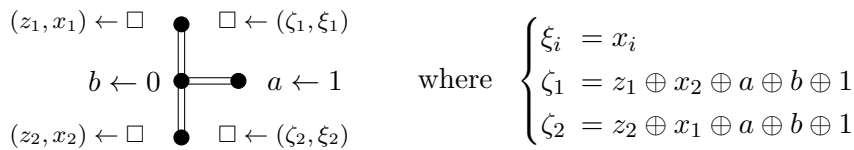
\begin{equation*}\label{eq:pat-CZ}
\begin{tikzpicture}
	\begin{pgfonlayer}{nodelayer}
		\node [style={small black dot}] (0) at (-3.75, 1.5) {};
		\node [style={small black dot}] (1) at (-3.75, -0) {};
		\node [style={small black dot}] (2) at (-2.25, -0) {};
		\node [style={small black dot}] (3) at (-3.75, -1.5) {};
		\node [style=none] (4) at (-0.5, -0) {$a\leftarrow 1$};
		\node [style=none] (5) at (-5.25, -0) {$b\leftarrow 0$};
		\node [style={left label}] (6) at (-4.75, 1.5) {$(z_1,x_1)\leftarrow \square$};
		\node [style={right label}] (7) at (-2.75, -1.5) {$\square \leftarrow (\zeta_2,\xi_2)$};
		\node [style={right label}] (8) at (-2.75, 1.5) {$\square \leftarrow (\zeta_1,\xi_1)$};
		\node [style={left label}] (9) at (-4.75, -1.5) {$(z_2,x_2)\leftarrow \square$};
	\end{pgfonlayer}
	\begin{pgfonlayer}{edgelayer}
		\draw [style={gs double edge}] (0) to (1);
		\draw [style={gs double edge}] (1) to (2);
		\draw [style={gs double edge}] (1) to (3);
	\end{pgfonlayer}
\end{tikzpicture} 
\qquad\textrm{where}\ \ \begin{cases}
\xi_i & \!\!\!= x_i \\
\zeta_1 & \!\!\!= z_1\oplus x_2\oplus a \oplus b \oplus 1 \\
\zeta_2 & \!\!\!= z_2\oplus x_1\oplus a \oplus b \oplus 1
\end{cases}
\end{equation*}
By conjugating one side with Hadamard gates we get a CNOT gate, and thus we can use this gate to achieve universality.

Note that the top and bottom qubits act as both inputs and outputs, so they are not measured. We measure $a$ in the Z-basis, and $b$ in the X-basis. Writing the resulting diagram out in ZX-calculus (again ignoring incoming Pauli errors for the moment) we get:
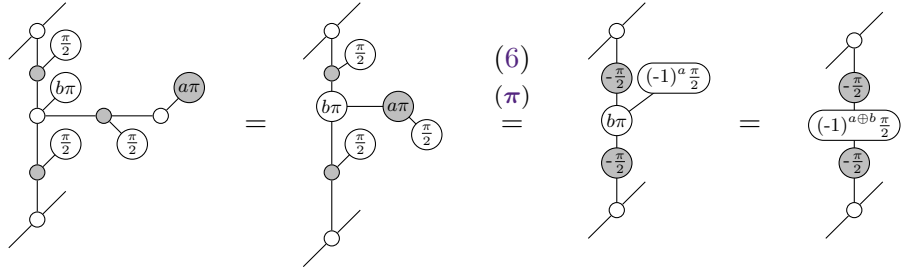
\ctikzfig{ms-zx-cnot}
Now we can use the conversion rule for Hadamards ($\begin{tikzpicture}
	\begin{pgfonlayer}{nodelayer}
		\node [style=none] (0) at (-0.25, 0) {};
		\node [style=none] (1) at (1.75, 0) {};
		\node [style=small hadamard] (2) at (0.75, 0) {};
		\node [style=none] (3) at (2.5, 0) {$=$};
		\node [style=none] (4) at (3.25, 0) {};
		\node [style=gray phase dot] (5) at (4, 0) {-$\frac\pi2$};
		\node [style=white phase dot] (6) at (5, 0) {-$\frac\pi2$};
		\node [style=gray phase dot] (7) at (6, 0) {-$\frac\pi2$};
		\node [style=none] (8) at (6.75, 0) {};
	\end{pgfonlayer}
	\begin{pgfonlayer}{edgelayer}
		\draw [style=simple] (0.center) to (2);
		\draw [style=simple] (2) to (1.center);
		\draw [style=simple] (4.center) to (8.center);
	\end{pgfonlayer}
\end{tikzpicture}$) on the middle part, giving:
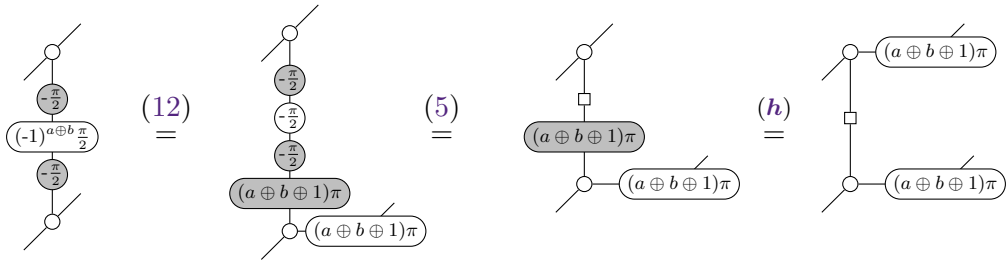
\ctikzfig{ms-zx-cnot2}
Pauli errors propagate through a CZ in the following way:
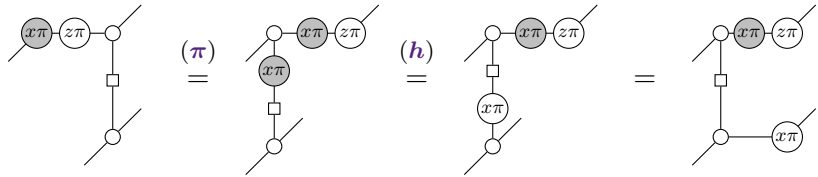
\ctikzfig{zx-cx-propagate}
and in exactly the same way for the other input. Putting the above derivation together with this error propagation gives the pattern fragment as specified above.

This pattern fragment implementing CZ has the additional property that if we measure $b$ in the Z basis instead of the X basis, it \emph{disconnects}. It doesn't matter in which basis we measure $a$, but let's take it to be the X basis. This pattern fragment is:
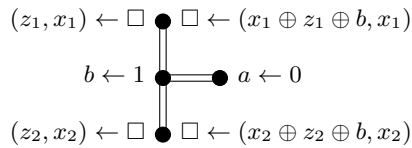
\begin{equation*}\label{eq:pat-SS}
\begin{tikzpicture}
	\begin{pgfonlayer}{nodelayer}
		\node [style=right label] (0) at (2, 0) {$a\leftarrow 0$};
		\node [style=right label] (1) at (0.5, -1.5) {$\square \leftarrow (x_2\oplus z_2\oplus b,x_2)$};
		\node [style=small black dot] (2) at (0, 0) {};
		\node [style=right label] (3) at (0.5, 1.5) {$\square \leftarrow (x_1\oplus z_1\oplus b,x_1)$};
		\node [style=left label] (4) at (-0.5, 1.5) {$(z_1,x_1)\leftarrow \square$};
		\node [style=small black dot] (5) at (1.5, 0) {};
		\node [style=small black dot] (6) at (0, 1.5) {};
		\node [style=left label] (7) at (-0.5, -1.5) {$(z_2,x_2)\leftarrow \square$};
		\node [style=small black dot] (8) at (0, -1.5) {};
		\node [style=left label] (9) at (-0.5, 0) {$b\leftarrow 1$};
	\end{pgfonlayer}
	\begin{pgfonlayer}{edgelayer}
		\draw [style=gs double edge] (6) to (2);
		\draw [style=gs double edge] (2) to (5);
		\draw [style=gs double edge] (2) to (8);
	\end{pgfonlayer}
\end{tikzpicture}
\end{equation*}

\noindent In ZX notation (ignoring incoming Pauli errors):
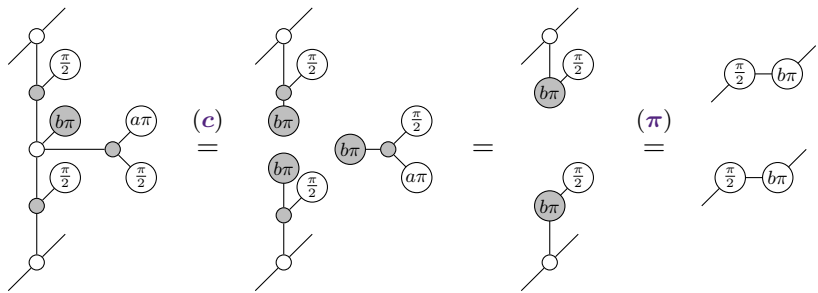
\ctikzfig{ms-zx-cnot-disconnect}
Hence, the choice of measurement basis for $b$ `switches' the CZ-gate on or off.

\section{Proof of universality}\label{sec:universal}

In the previous section,  we have constructed a pattern fragment that can implement a $T$ gate and an $H$ gate depending on the chosen measurements. Combining these, we can produce $H$ and $T$, which suffice to  approximate any single-qubit unitary. Combining this with the entangling gate we demonstrated at the end of the previous section suffices for universality.

It only remains to combine these patterns into a configuration that allows us to combine them arbitrarily. We will construct a fragment that is a combination of the simple blocks described above which fits neatly into a 2D square lattice. If we simply compose two of the `E'-shaped blocks from the previous section with a CZ block rotated 90 degrees we get:
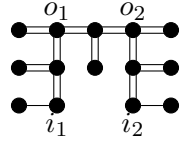
\ctikzfig{ms-brick-minimal}
where the $i$'s denote inputs and $o$'s denote outputs.

These bricks however don't fit together in a square lattice, since there is no useful tiling we can produce without some qubits overlapping. However, we can solve this problem by considering a slightly larger brick, where the E-shape on the right is offset downward and extra double-edges are added to $i_1, i_2$ and $o_2$:
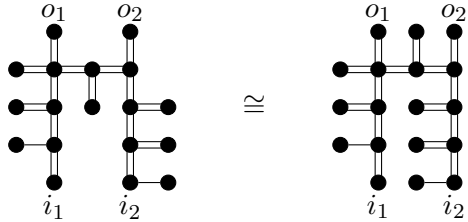
\[ \begin{tikzpicture}
	\begin{pgfonlayer}{nodelayer}
		\node [style=small black dot] (0) at (-3, 1) {};
		\node [style=small black dot] (1) at (-2, 1) {};
		\node [style=small black dot] (2) at (-1, 1) {};
		\node [style=small black dot] (3) at (-3, 0) {};
		\node [style=small black dot] (4) at (-2, 0) {};
		\node [style=small black dot] (5) at (-3, -1) {};
		\node [style=small black dot] (6) at (-1, 0) {};
		\node [style=small black dot] (7) at (-1, -1) {};
		\node [style=small black dot] (8) at (0, -1) {};
		\node [style=small black dot] (9) at (-4, -1) {};
		\node [style=small black dot] (10) at (0, -2) {};
		\node [style=small black dot] (11) at (0, 0) {};
		\node [style=small black dot] (12) at (-4, 1) {};
		\node [style=small black dot] (13) at (-3, -2) {};
		\node [style=small black dot] (14) at (-1, -2) {};
		\node [style=small black dot] (15) at (-1, 2) {};
		\node [style=none] (16) at (-3, -2.65) {$i_1$};
		\node [style=none] (17) at (-1, -2.65) {$i_2$};
		\node [style=none] (18) at (-3, 2.5) {$o_1$};
		\node [style=none] (19) at (-1, 2.5) {$o_2$};
		\node [style=small black dot] (20) at (-3, 2) {};
		\node [style=small black dot] (21) at (-4, 0) {};
	\end{pgfonlayer}
	\begin{pgfonlayer}{edgelayer}
		\draw [style=gs double edge] (0) to (1);
		\draw [style=gs double edge] (1) to (4);
		\draw [style=gs double edge] (1) to (2);
		\draw [style=gs double edge] (0) to (3);
		\draw [style=gs double edge] (3) to (5);
		\draw [style=gs double edge] (2) to (6);
		\draw [style=gs double edge] (6) to (7);
		\draw [style=gs double edge] (0) to (12);
		\draw [style=gs double edge] (6) to (11);
		\draw [style=gs edge] (9) to (5);
		\draw [style=gs double edge] (5) to (13);
		\draw [style=gs double edge] (14) to (7);
		\draw [style=gs double edge] (2) to (15);
		\draw [style=gs edge] (14) to (10);
		\draw [style=gs double edge] (7) to (8);
		\draw [style=gs double edge] (3) to (21);
		\draw [style=gs double edge] (0) to (20);
	\end{pgfonlayer}
\end{tikzpicture} \qquad \cong \qquad  \begin{tikzpicture}
	\begin{pgfonlayer}{nodelayer}
		\node [style=small black dot] (0) at (-3, 1) {};
		\node [style=small black dot] (1) at (-2, 1) {};
		\node [style=small black dot] (2) at (-1, 1) {};
		\node [style=small black dot] (3) at (-3, 0) {};
		\node [style=small black dot] (4) at (-2, 2) {};
		\node [style=small black dot] (5) at (-3, -1) {};
		\node [style=small black dot] (6) at (-1, 0) {};
		\node [style=small black dot] (7) at (-1, -1) {};
		\node [style=small black dot] (8) at (-2, -1) {};
		\node [style=small black dot] (9) at (-4, -1) {};
		\node [style=small black dot] (10) at (-2, -2) {};
		\node [style=small black dot] (11) at (-2, 0) {};
		\node [style=small black dot] (12) at (-4, 1) {};
		\node [style=small black dot] (13) at (-3, -2) {};
		\node [style=small black dot] (14) at (-1, -2) {};
		\node [style=small black dot] (15) at (-1, 2) {};
		\node [style=none] (16) at (-3, -2.625) {$i_1$};
		\node [style=none] (17) at (-1, -2.625) {$i_2$};
		\node [style=none] (18) at (-3, 2.5) {$o_1$};
		\node [style=none] (19) at (-1, 2.5) {$o_2$};
		\node [style=small black dot] (21) at (-4, 0) {};
		\node [style=small black dot] (22) at (-3, 2) {};
	\end{pgfonlayer}
	\begin{pgfonlayer}{edgelayer}
		\draw [style=gs double edge] (0) to (1);
		\draw [style=gs double edge] (1) to (4);
		\draw [style=gs double edge] (1) to (2);
		\draw [style=gs double edge] (0) to (3);
		\draw [style=gs double edge] (3) to (5);
		\draw [style=gs double edge] (2) to (6);
		\draw [style=gs double edge] (6) to (7);
		\draw [style=gs double edge] (0) to (12);
		\draw [style=gs double edge] (6) to (11);
		\draw [style=gs edge] (9) to (5);
		\draw [style=gs double edge] (5) to (13);
		\draw [style=gs double edge] (14) to (7);
		\draw [style=gs double edge] (2) to (15);
		\draw [style=gs edge] (14) to (10);
		\draw [style=gs double edge] (7) to (8);
		\draw [style=gs double edge] (3) to (21);
		\draw [style=gs double edge] (0) to (22);
	\end{pgfonlayer}
\end{tikzpicture} \]
We can picture the paths from $i_1$ to $o_1$ and $i_2$ to $o_2$ as two qubits passing through a circuit. The E-shapes on the left and the right can be used to apply $S$, $T$, or $H$ gates depending on the choice of pattern. Similarly, the shape connecting the two qubits can be used to apply CZ or $S \otimes S$ to both qubits. The extra edges will always introduce $HSH$ gates. By selecting these patterns, we can implement $3 \cdot 3 \cdot 2$ different two-qubit unitaries $U$. When these are arranged as follows:



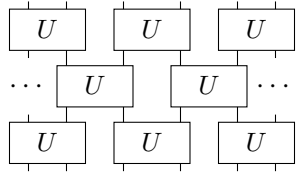
\ctikzfig{unitary-brick}
it is straightforward to verify that we can implement any Clifford+T circuit. Hence, we obtain
a model allowing for universal quantum computation, much like in Ref.~\cite{brickworkuniversal}.

The asymmetry present in the brick allows us to efficiently tile them:
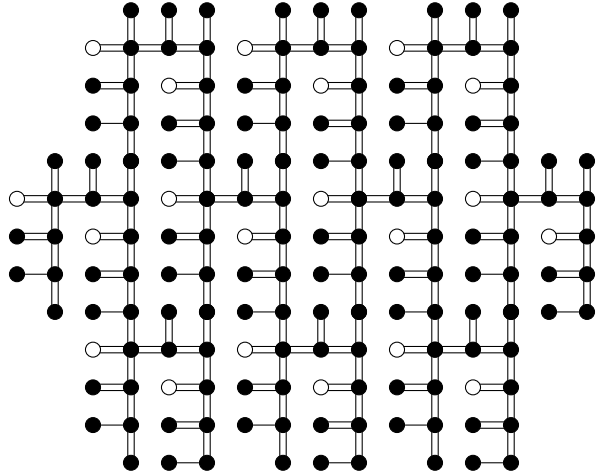
\ctikzfig{ms-brickwork}
We see that all locations in a square grid are used. Qubits whose measurement could potentially depend on prior outcomes are shown in white above. Of those, only the corrections forming part of the T pattern described rely on feed-forward. Hence \textit{all} of the other qubits can be measured simultaneously at the beginning of the computation.

This graph state can be viewed as consisting of lanes which carry the computation forward, attached to which are `hairs' which introduce extra phases. `Shaving off' all of the hairs reveals a state that has a somewhat similar structure to the brickwork state from Ref.~\cite{brickworkuniversal}:
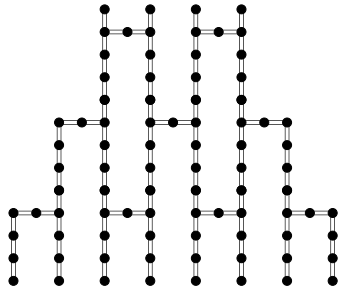
\begin{center}
  \scalebox{0.6}{\begin{tikzpicture}
	\begin{pgfonlayer}{nodelayer}
		\node [style={small black dot}] (0) at (-3, 1) {};
		\node [style={small black dot}] (1) at (-2, 1) {};
		\node [style={small black dot}] (2) at (-1, 1) {};
		\node [style={small black dot}] (3) at (-3, -0) {};
		\node [style={small black dot}] (4) at (-3, -1) {};
		\node [style={small black dot}] (5) at (-1, -0) {};
		\node [style={small black dot}] (6) at (-1, -1) {};
		\node [style={small black dot}] (7) at (-3, -2) {};
		\node [style={small black dot}] (8) at (-1, -2) {};
		\node [style={small black dot}] (9) at (-3, 2) {};
		\node [style={small black dot}] (10) at (-1, 2) {};
		\node [style={small black dot}] (11) at (-3, 6) {};
		\node [style={small black dot}] (12) at (-5, 3) {};
		\node [style={small black dot}] (13) at (-3, 3) {};
		\node [style={small black dot}] (14) at (-3, 5) {};
		\node [style={small black dot}] (15) at (-5, 6) {};
		\node [style={small black dot}] (16) at (-3, 4) {};
		\node [style={small black dot}] (17) at (-5, 2) {};
		\node [style={small black dot}] (18) at (-5, 4) {};
		\node [style={small black dot}] (19) at (-3, 2) {};
		\node [style={small black dot}] (20) at (-4, 5) {};
		\node [style={small black dot}] (21) at (-5, 5) {};
		\node [style={small black dot}] (22) at (1, 6) {};
		\node [style={small black dot}] (23) at (1, 3) {};
		\node [style={small black dot}] (24) at (1, 5) {};
		\node [style={small black dot}] (25) at (0, 5) {};
		\node [style={small black dot}] (26) at (-1, 2) {};
		\node [style={small black dot}] (27) at (-1, 5) {};
		\node [style={small black dot}] (28) at (-1, 6) {};
		\node [style={small black dot}] (29) at (1, 4) {};
		\node [style={small black dot}] (30) at (1, 2) {};
		\node [style={small black dot}] (31) at (-1, 3) {};
		\node [style={small black dot}] (32) at (-1, 4) {};
		\node [style={small black dot}] (33) at (1, -0) {};
		\node [style={small black dot}] (34) at (3, 1) {};
		\node [style={small black dot}] (35) at (1, -2) {};
		\node [style={small black dot}] (36) at (3, -0) {};
		\node [style={small black dot}] (37) at (3, -1) {};
		\node [style={small black dot}] (38) at (2, 1) {};
		\node [style={small black dot}] (39) at (1, 1) {};
		\node [style={small black dot}] (40) at (1, 2) {};
		\node [style={small black dot}] (41) at (3, -2) {};
		\node [style={small black dot}] (42) at (1, -1) {};
		\node [style={small black dot}] (43) at (-5, -4) {};
		\node [style={small black dot}] (44) at (-3, -4) {};
		\node [style={small black dot}] (45) at (-5, -3) {};
		\node [style={small black dot}] (46) at (-5, -5) {};
		\node [style={small black dot}] (47) at (-3, -5) {};
		\node [style={small black dot}] (48) at (-4, -3) {};
		\node [style={small black dot}] (49) at (-5, -6) {};
		\node [style={small black dot}] (50) at (-5, -2) {};
		\node [style={small black dot}] (51) at (-3, -3) {};
		\node [style={small black dot}] (52) at (-3, -6) {};
		\node [style={small black dot}] (53) at (-3, -2) {};
		\node [style={small black dot}] (54) at (-5, -0) {};
		\node [style={small black dot}] (55) at (-5, -2) {};
		\node [style={small black dot}] (56) at (-5, 2) {};
		\node [style={small black dot}] (57) at (-7, 1) {};
		\node [style={small black dot}] (58) at (-7, -2) {};
		\node [style={small black dot}] (59) at (-7, -0) {};
		\node [style={small black dot}] (60) at (-6, 1) {};
		\node [style={small black dot}] (61) at (-7, -1) {};
		\node [style={small black dot}] (62) at (-5, -1) {};
		\node [style={small black dot}] (63) at (-5, 1) {};
		\node [style={small black dot}] (64) at (1, -4) {};
		\node [style={small black dot}] (65) at (-1, -5) {};
		\node [style={small black dot}] (66) at (-1, -4) {};
		\node [style={small black dot}] (67) at (1, -3) {};
		\node [style={small black dot}] (68) at (1, -6) {};
		\node [style={small black dot}] (69) at (1, -2) {};
		\node [style={small black dot}] (70) at (-1, -3) {};
		\node [style={small black dot}] (71) at (0, -3) {};
		\node [style={small black dot}] (72) at (1, -5) {};
		\node [style={small black dot}] (73) at (-1, -2) {};
		\node [style={small black dot}] (74) at (-1, -6) {};
		\node [style={small black dot}] (75) at (3, -5) {};
		\node [style={small black dot}] (76) at (3, -4) {};
		\node [style={small black dot}] (77) at (3, -3) {};
		\node [style={small black dot}] (78) at (3, -2) {};
		\node [style={small black dot}] (79) at (3, -6) {};
		\node [style={small black dot}] (80) at (5, -3) {};
		\node [style={small black dot}] (81) at (4, -3) {};
		\node [style={small black dot}] (82) at (-9, -6) {};
		\node [style={small black dot}] (83) at (-7, -3) {};
		\node [style={small black dot}] (84) at (-9, -4) {};
		\node [style={small black dot}] (85) at (-7, -4) {};
		\node [style={small black dot}] (86) at (-8, -3) {};
		\node [style={small black dot}] (87) at (-7, -2) {};
		\node [style={small black dot}] (88) at (-9, -3) {};
		\node [style={small black dot}] (89) at (-7, -5) {};
		\node [style={small black dot}] (90) at (-9, -5) {};
		\node [style={small black dot}] (91) at (-7, -6) {};
		\node [style={small black dot}] (92) at (5, -5) {};
		\node [style={small black dot}] (93) at (5, -4) {};
		\node [style={small black dot}] (94) at (5, -6) {};
	\end{pgfonlayer}
	\begin{pgfonlayer}{edgelayer}
		\draw [style={gs double edge}] (0) to (1);
		\draw [style={gs double edge}] (1) to (2);
		\draw [style={gs double edge}] (0) to (3);
		\draw [style={gs double edge}] (3) to (4);
		\draw [style={gs double edge}] (2) to (5);
		\draw [style={gs double edge}] (5) to (6);
		\draw [style={gs double edge}] (4) to (7);
		\draw [style={gs double edge}] (8) to (6);
		\draw [style={gs double edge}] (0) to (9);
		\draw [style={gs double edge}] (2) to (10);
		\draw [style={gs double edge}] (21) to (20);
		\draw [style={gs double edge}] (20) to (14);
		\draw [style={gs double edge}] (21) to (18);
		\draw [style={gs double edge}] (18) to (12);
		\draw [style={gs double edge}] (14) to (16);
		\draw [style={gs double edge}] (16) to (13);
		\draw [style={gs double edge}] (12) to (17);
		\draw [style={gs double edge}] (19) to (13);
		\draw [style={gs double edge}] (21) to (15);
		\draw [style={gs double edge}] (14) to (11);
		\draw [style={gs double edge}] (27) to (25);
		\draw [style={gs double edge}] (25) to (24);
		\draw [style={gs double edge}] (27) to (32);
		\draw [style={gs double edge}] (32) to (31);
		\draw [style={gs double edge}] (24) to (29);
		\draw [style={gs double edge}] (29) to (23);
		\draw [style={gs double edge}] (31) to (26);
		\draw [style={gs double edge}] (30) to (23);
		\draw [style={gs double edge}] (27) to (28);
		\draw [style={gs double edge}] (24) to (22);
		\draw [style={gs double edge}] (39) to (38);
		\draw [style={gs double edge}] (38) to (34);
		\draw [style={gs double edge}] (39) to (33);
		\draw [style={gs double edge}] (33) to (42);
		\draw [style={gs double edge}] (34) to (36);
		\draw [style={gs double edge}] (36) to (37);
		\draw [style={gs double edge}] (42) to (35);
		\draw [style={gs double edge}] (41) to (37);
		\draw [style={gs double edge}] (39) to (40);
		\draw [style={gs double edge}] (45) to (48);
		\draw [style={gs double edge}] (48) to (51);
		\draw [style={gs double edge}] (45) to (43);
		\draw [style={gs double edge}] (43) to (46);
		\draw [style={gs double edge}] (51) to (44);
		\draw [style={gs double edge}] (44) to (47);
		\draw [style={gs double edge}] (46) to (49);
		\draw [style={gs double edge}] (52) to (47);
		\draw [style={gs double edge}] (45) to (50);
		\draw [style={gs double edge}] (51) to (53);
		\draw [style={gs double edge}] (57) to (60);
		\draw [style={gs double edge}] (60) to (63);
		\draw [style={gs double edge}] (57) to (59);
		\draw [style={gs double edge}] (59) to (61);
		\draw [style={gs double edge}] (63) to (54);
		\draw [style={gs double edge}] (54) to (62);
		\draw [style={gs double edge}] (61) to (58);
		\draw [style={gs double edge}] (55) to (62);
		\draw [style={gs double edge}] (63) to (56);
		\draw [style={gs double edge}] (70) to (71);
		\draw [style={gs double edge}] (71) to (67);
		\draw [style={gs double edge}] (70) to (66);
		\draw [style={gs double edge}] (66) to (65);
		\draw [style={gs double edge}] (67) to (64);
		\draw [style={gs double edge}] (64) to (72);
		\draw [style={gs double edge}] (65) to (74);
		\draw [style={gs double edge}] (68) to (72);
		\draw [style={gs double edge}] (70) to (73);
		\draw [style={gs double edge}] (67) to (69);
		\draw [style={gs double edge}] (77) to (81);
		\draw [style={gs double edge}] (81) to (80);
		\draw [style={gs double edge}] (77) to (76);
		\draw [style={gs double edge}] (76) to (75);
		\draw [style={gs double edge}] (75) to (79);
		\draw [style={gs double edge}] (77) to (78);
		\draw [style={gs double edge}] (88) to (86);
		\draw [style={gs double edge}] (86) to (83);
		\draw [style={gs double edge}] (88) to (84);
		\draw [style={gs double edge}] (84) to (90);
		\draw [style={gs double edge}] (83) to (85);
		\draw [style={gs double edge}] (85) to (89);
		\draw [style={gs double edge}] (90) to (82);
		\draw [style={gs double edge}] (91) to (89);
		\draw [style={gs double edge}] (83) to (87);
		\draw [style={gs double edge}] (94) to (92);
		\draw [style={gs double edge}] (93) to (92);
		\draw [style={gs double edge}] (80) to (93);
	\end{pgfonlayer}
\end{tikzpicture}}
\end{center}

While our primitive computational `brick' doesn't seem to be particularly canonical, the fact that it is missing some edges from a square lattice could have advantages when thinking about space-limited architectures. For instance, after dropping the extra `dummy' qubits $i_1$ and $o_1$ and a bit of folding, the resulting 16-qubit pattern fits into the 17-qubit `ninja star' design of the superconducting chip proposed in~\cite{superconductingdelft}, which is designed primarily for implementing a 17-qubit surface code:
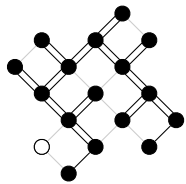
\begin{center}
\rotatebox{45}{\begin{tikzpicture}
	\begin{pgfonlayer}{nodelayer}
		\node [style=small black dot] (0) at (-3, 1) {};
		\node [style=small black dot] (1) at (-2, 1) {};
		\node [style=small black dot] (2) at (-1, 1) {};
		\node [style=small black dot] (3) at (-3, 0) {};
		\node [style=small black dot] (4) at (-2, 2) {};
		\node [style=small black dot] (5) at (-3, -1) {};
		\node [style=small black dot] (6) at (-1, 0) {};
		\node [style=small black dot] (7) at (-1, -1) {};
		\node [style=small black dot] (8) at (-2, -1) {};
		\node [style=small black dot] (9) at (-4, -1) {};
		\node [style=small black dot] (10) at (-2, -2) {};
		\node [style=small black dot] (11) at (0, 0) {};
		\node [style=small black dot] (12) at (-3, 2) {};
		\node [style=small black dot] (13) at (-1, -2) {};
		\node [style=small black dot] (14) at (0, 1) {};
		\node [style=small black dot] (15) at (-2, 0) {};
		\node [style=mbqc input dot] (16) at (-4, 0) {};
		\node [style=mbqc input dot] (17) at (-4, 0) {};
	\end{pgfonlayer}
	\begin{pgfonlayer}{edgelayer}
		\draw [style=gs double edge] (0) to (1);
		\draw [style=gs double edge] (1) to (4);
		\draw [style=gs double edge] (1) to (2);
		\draw [style=gs double edge] (0) to (3);
		\draw [style=gs double edge] (3) to (5);
		\draw [style=gs double edge] (2) to (6);
		\draw [style=gs double edge] (6) to (7);
		\draw [style=gs double edge] (0) to (12);
		\draw [style=gs double edge] (6) to (11);
		\draw [style=gs edge] (9) to (5);
		\draw [style=gs double edge] (13) to (7);
		\draw [style=gs double edge] (2) to (14);
		\draw [style=gs edge] (13) to (10);
		\draw [style=gs double edge] (7) to (8);
		\draw [style=gs double edge] (3) to (15);
		\draw [style=gray edge] (9) to (16);
		\draw [style=gray edge] (16) to (3);
		\draw [style=gray edge] (15) to (8);
		\draw [style=gray edge] (8) to (5);
		\draw [style=gray edge] (8) to (10);
		\draw [style=gray edge] (15) to (6);
		\draw [style=gray edge] (15) to (1);
		\draw [style=gray edge] (12) to (4);
		\draw [style=gray edge] (11) to (14);
	\end{pgfonlayer}
\end{tikzpicture}}
\end{center}
This means that a proof of concept for this computational model is potentially close at hand. Looking at the way the `E'-shape implements the T-gate we also see that the middle `hair' is actually not necessary. Removing this qubit allows us to fit a universal brick inside the superconducting chip of Ref.~\cite{otterbach2017unsupervised}.

\section{Climbing the Clifford hierarchy}\label{sec:clifhier}

The construction of a deterministic feed-forward strategy in the previous sections relies on the fact that a sign error in a T gate, i.e.\ a $\pi/4$ rotation, can be corrected by applying a $\pi/2$ rotation, and then correcting a sign error in a $\pi/2$ rotation by selectively applying a $\pi$ rotation. Since such a $\pi$ rotation can be commuted past all other gates, the resulting error can be handled at the end of the computation in a classical manner. 

We will now show that this works not only for $\pi/4$, but for all angles $\pi/2^n$ where $2 \leq n \in \mathbb N$. While the P-graphs of the previous sections had a single edge represent a $P(\frac\pi4)$ gate, we will now let a single edge represent $P(\frac{\pi}{2^n})$. If we have $k$ edges between two vertices then this represents a $P(\frac{\pi}{2^n})^k = P(\frac{k\pi}{2^n})$ gate. Now consider the following fragment:

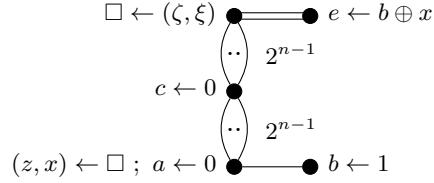
\ctikzfig{ms-hierarchy}
Here the $2^{n-1}$ refers to the amount of wires between the vertices. I.e.\ this means that there will be a $P(\pi/2)$ gate between the qubits there. Let us simplify the corresponding diagram in the ZX-calculus. We will not yet apply the measurement of the qubit $e$.
\ctikzfig{ms-hierarchy-proof1}

If $b=0$, then we have the rotation we want, so we can measure the remaining qubit $e$ in the $X$ basis, and we see that we are left with a $\pi/2^n$ rotation with some Pauli error. If $b=1$, however, we will measure $e$ in the Z-basis, and we calculate:

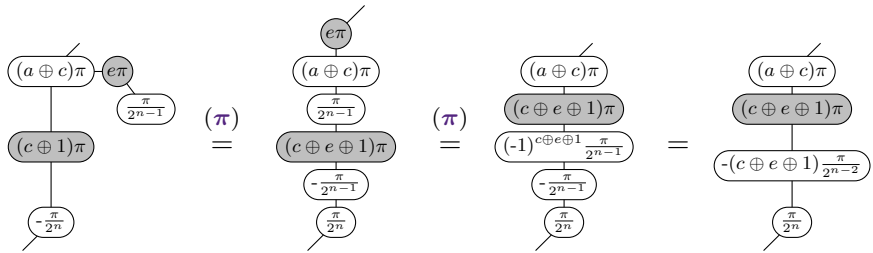
\ctikzfig{ms-hierarchy-proof2}

If $c\oplus e = 1$, then this is the desired computation. Otherwise, we are left with an unwanted $\pi/2^{n-2}$ rotation. Since this undesired rotation is heralded by the outcome of a measurement, we can however decide to do a $\pi/2^{n-2}$ in its future to cancel out this rotation using exactly the same procedure. Trying to do this extra $\pi/2^{n-2}$ rotation could then introduce an unwanted $\pi/2^{n-3}$ rotation. After $n-2$ repetitions of this protocol we are therefore left with a $\pi$ error that can safely be incorporated into the classical feed-forward.

The relevant concept to understand this kind of iteration of rotations is that of the \emph{Clifford hierarchy}. The first level $\mathcal{C}_1$ is defined to be the set of tensor products of the identity and the Pauli unitaries. 
The higher levels are then iteratively defined to be the multiqubit unitaries that send the Pauli unitaries to a lower level of the hierarchy: $\mathcal{C}_n := \{U ~|~ \forall V\in \mathcal{C}_1.\  UVU^\dagger \in \mathcal{C}_{n-1}\}$. 
With this definition we can see that $\mathcal{C}_2$ consists of exactly the usual Clifford operators. While $\mathcal{C}_1$ and $\mathcal{C}_2$ are closed under composition, the higher levels no longer form groups. 
In fact, not much is known about the general structure of the higher sets in the Clifford hierarchy. 
What we do know however is that the \emph{diagonal} unitaries in $\mathcal{C}_n$ always form a group, and that for $n\geq 2$ each of these diagonal elements can be constructed using Clifford operations and the $\pi/2^{n-1}$ $Z$ rotation~\cite{cui2017diagonal}.
Using the above description of a deterministic implementation of a $\pi/2^{n-1}$ rotation we have therefore found a deterministic measurement-based model that can implement any diagonal $n$-th level Clifford operation using just Pauli measurements.

\section{Conclusions and further work}

We presented a novel family of graph states that lead to approximately universal quantum computation using just measurements in two bases. Furthermore, depending on the chosen parameter, diagonal gates of arbitrarily high levels of the Clifford hierarchy can be implemented.


In the scheme we gave, each of the qubits whose measurement depends on previous outcomes is a correction involved in implementing a single $T$-gate. Hence, the number of required measurement/feed-forward cycles is clearly related to the $T$-depth of the associated circuit. We intend to make this relationship precise and relate to similar results known for the one-way model concerning pattern depth and parallelisation of measurements. Another avenue of future work is to understand the general properties of feed-forward in P-graph states. For instance, concepts like Flow and gFlow \cite{GFlow,MarkhamKashefi} are likely to apply with little modification to the `double edge' portion of a P-graph state, so it will be interesting to see if this can be extended.

This work also highlights a link between the form of a parity-phase gate derived in equation \eqref{eq:virtual-qubit} and quantum computing with magic states. It may be useful to consider if the representation of Ising-type interactions (i.e. parity-phase gates) as `virtual' magic states can be exploited for magic state distillation. For instance, on ion trap based architectures, it is possible to introduce $O(n^2)$ parity-phase gates in a single time step using an $n$-qubit interaction \cite{molmersorensen1999}. It is a topic of future work to try to make use of this curious property in the construction of fault-tolerant protocols.


The interactions needed to make the graph states described in this paper are available `natively' in both ion trap and superconducting quantum computing hardware which means proofs of concept could be implemented in a short timeframe. However, given long turnaround times for quantum measurements and feed-forward, it remains unclear if such a measurement-based scheme would yield benefits over the circuit model on such architectures. On the other hand, measurement-based schemes have already had some success in quantum optics~\cite{walther2005experimental}, where deterministic application of multi-qubit gates remains a significant challenge. While the scheme we gave relied on resource states with a very specific structure, its likely that this could be relaxed using techniques similar to those employed in producing perfect cluster states from imperfect lattices (e.g.~\cite{morley2017physical}). Furthermore, the use of multiple kinds of edges between qubits creates a possibility for more multiple successful outcomes for non-deterministic entangling operations. That is, \textit{known} errors giving rise to non-maximal entanglement between pairs of qubits could still yield good resource states for universal deterministic computation. This could, for example, be exploited in models of universal quantum computation using linear optical devices and non-deterministic fusion gates~\cite{gimeno2015three}.

Another advantage for our approach is the availability of automated tools for reasoning with, and transforming ZX-diagrams. The graphical proof assistant Quantomatic~\cite{kissinger2015quantomatic} makes it possible to scale calculations in the style of this paper to large and more elaborate pattern fragments, and the quantum compilation tool PyZX~\cite{pyzx,pyzxtheory}---which uses the ZX-calculus as a native intermediate representation for computations---can be straightforwardly adapted to produce ZX-based measurement patterns as a target architecture.

\textbf{Acknowledgements}: This work is supported by the ERC under the European Union’s Seventh Framework Programme (FP7/2007-2013) / ERC grant n$^\text{o}$ 320571 and AFOSR grant FA2386-18-1-4028. The authors would like to thank Brian Tarasinksi and Martin Sepiol for useful discussions about superconducting qubits and the M\o{}lmer-S\o{}rensen interaction in ion traps, respectively.

\bibliographystyle{plainnat}
\bibliography{main}

\end{document}